\documentclass[acmlarge]{acmart}
\usepackage{multirow}
\usepackage{subcaption}
\usepackage{graphicx}

\usepackage{booktabs} % For formal tables

\usepackage[ruled]{algorithm2e} % For algorithms

\SetAlFnt{\small}
\SetAlCapFnt{\small}
\SetAlCapNameFnt{\small}
\SetAlCapHSkip{0pt}
\IncMargin{-\parindent}

\received{May 2017}
\received[accepted]{June 2017}

\begin{document}
\title{Detecting Gaze Towards Eyes in Natural Social Interactions and Its Use in Child Assessment} 
\author{Eunji Chong}
\author{Katha Chanda}
\affiliation{%
  \institution{Georgia Institute of Technology}
  \department{School of Interactive Computing}
  \city{Atlanta}
  \state{GA}
  \country{USA}}

\author{Zhefan Ye}
\affiliation{%
  \institution{University of Michigan}
  \department{Computer Science and Engineering}
  \city{Ann Arbor}
  \state{MI}
  \country{USA}}
  
\author{Audrey Southerland}
\author{Nataniel Ruiz}
\affiliation{%
  \institution{Georgia Institute of Technology}
  \department{School of Interactive Computing}
  \city{Atlanta}
  \state{GA}
  \country{USA}}
  
\author{Rebecca M. Jones}
\affiliation{%
  \institution{Weill Cornell Medicine}
  \city{White Plains}
  \state{NY}
  \country{USA}}
  
\author{Agata Rozga$^*$}
\author{James M. Rehg$^*$}
\affiliation{%
  \institution{Georgia Institute of Technology}
  \department{School of Interactive Computing}
  \city{Atlanta}
  \state{GA}
  \country{USA}}  

\begin{abstract}
Eye contact is a crucial element of non-verbal communication that signifies interest, attention, and participation in social interactions. As a result, measures of eye contact arise in a variety of applications such as the assessment of the social communication skills of children at risk for developmental disorders such as autism, or the analysis of turn-taking and social roles during group meetings. However, the automated measurement of visual attention during naturalistic social interactions is challenging due to the difficulty of estimating a subject's looking direction from video. This paper proposes a novel approach to eye contact detection during adult-child social interactions in which the adult wears a point-of-view camera which captures an egocentric view of the child's behavior. By analyzing the child's face regions and inferring their head pose we can accurately identify the onset and duration of the child's looks to their social partner's eyes. We introduce the Pose-Implicit CNN, a novel deep learning architecture that predicts eye contact while implicitly estimating the head pose. We present a fully automated system for eye contact detection that solves the sub-problems of end-to-end feature learning and pose estimation using deep neural networks. To train our models, we use a dataset comprising 22 hours of 156 play session videos from over 100 children, half of whom are diagnosed with Autism Spectrum Disorder. We report an overall precision of 0.76, recall of 0.80, and an area under the precision-recall curve of 0.79, all of which are significant improvements over existing methods.
\end{abstract}

\setcopyright{acmcopyright}
\acmJournal{IMWUT}
\acmYear{2017} \acmVolume{1} \acmNumber{3} \acmArticle{43} \acmMonth{9} \acmPrice{}\acmDOI{10.1145/3131902}




\thanks{$^*$ Rozga and Rehg share senior authorship of this work. 
This study was funded in part by the Simons Foundation under grants 336363 and 383667, as well as the National Science Foundation under grant IIS-1029679.

  Authors' addresses: E. Chong, K. Chanda, A. Southerland, Nataniel Ruiz, A. Rozga {and} J. M. Rehg, Center for Behavioral Imaging and School of Interactive Computing, College of Computing, Georgia Institute of Technology; Z. Ye, Computer Science and Engineering, University of Michigan; R. M. Jones, Weill Cornell Medicine, Center for Autism and the Developing Brain.}

\maketitle

\renewcommand{\shortauthors}{E. Chong et al.}

\section{Introduction}
Eye contact is one of the most basic and powerful forms of nonverbal communication that humans use from the first months of life~\cite{brazelton1975early}. It plays a crucial role in social interactions where it is used for various purposes: to express interest and attentiveness, to signal a wish to participate, and to regulate interactions~\cite{argyle1965eye,kleinke1986gaze}. Moreover, eye contact is a key constituting element in joint attention, in which gaze and gestures are used to spontaneously create or indicate a shared point of reference with another person~\cite{mundy2006joint}. Individuals with autism spectrum disorder (ASD), a group of developmental disorders characterized by difficulties in engaging with the social world, show atypical patterns of gaze, eye contact, and joint attention; they look less at the eyes of others and respond less to the calling of one's name than their typically developing (TD) peers. These patterns have been identified as among the earliest indicators of autism in the first two years of life~\cite{hutman2012selective,rozga2011behavioral,jones2008absence}, and continue to remain distinctive throughout childhood and adolescence~\cite{sigman1998emanuel,klin2002visual}.

\begin{figure}[t]
  \centering
    \begin{subfigure}[t]{0.18\textwidth}
        \includegraphics[width=1\textwidth]{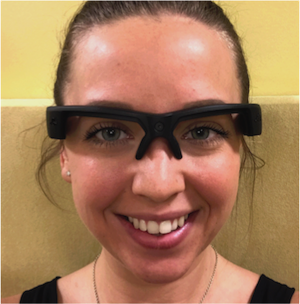}
        \caption{Examiner's face wearing the recording glasses.}\label{fig:setup1}
    \end{subfigure}
    \begin{subfigure}[t]{0.5\textwidth}
        \includegraphics[width=1\textwidth]{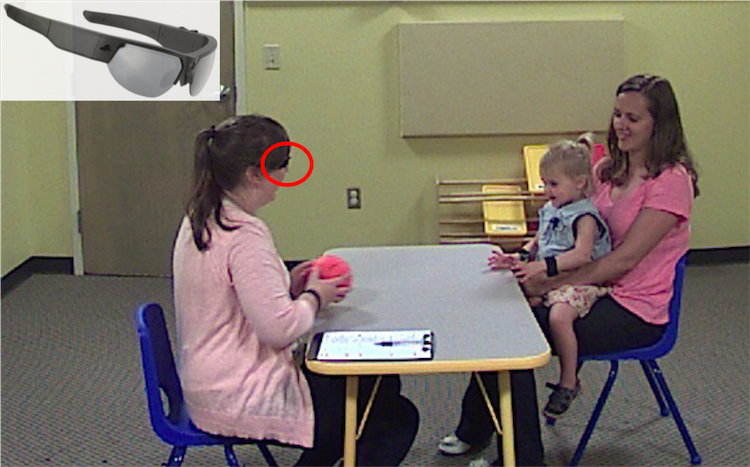}
        \caption{Child and examiner interact face-to-face.}
    \end{subfigure}
    \begin{subfigure}[t]{0.21\textwidth}
        \includegraphics[width=1\textwidth]{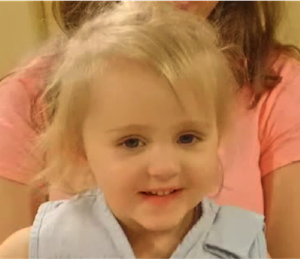}
        \caption{Child's face captured by the wearable camera.}
    \end{subfigure}
  \caption{\textbf{Our data collection setup.} Examiner wears a camera embedded in the bridge over the nose, thus the camera is naturally aligned with the adult's eyes and when the child looks to the adult's eyes it will be captured as a look towards the camera.}
\end{figure}
ASD affects a large population of children: 1 in 68 children in the United States have ASD and approximately $1-2\%$ of children are affected worldwide~\cite{cdctable}. While the cause of ASD is unknown, the consensus among experts is that early diagnosis and intervention can substantially increase positive outcomes~\cite{daniels2014explaining}. The development of automated measures for the early behavioral signs of autism (e.g., lack of eye contact) would facilitate large-scale screening and decrease the age of diagnosis. In addition, the continuous measurement of social behaviors such as eye contact in children who are receiving treatment would be extremely valuable in tailoring interventions and quantifying their effectiveness.

In spite of its importance as a measure of social communication, there exist very few  methods for measuring eye contact during naturalistic social interactions. The only widely-utilized paradigm occurs in the context of psychology research studies, in which social interactions are recorded from one or more cameras and manual annotation by multiple human raters provides a measure of eye contact. This is an inherently subjective determination and it clearly does not scale to broad use in screening and outcome assessment.

Eye tracking technology provides another widely-used approach for automatically assessing gaze behavior in social contexts. However, it imposes substantial constraints on the interaction, requiring a child to either passively view content on a monitor screen or wear head-mounted eye tracking hardware. Neither of these scenarios are appropriate for our target use case of naturalistic, face-to-face interactions. Since face-to-face interactions between an adult and a child are the basis for all screening, diagnosis, treatment, and assessment, it is a critical use case. While eye tracking technology is likely to soon be available on a large scale through its integration into tablets and laptop screens, it is unclear to what extent a child's gaze behavior in such settings reflects their gaze behavior with real-life social partners~\cite{foulsham2011and}.

To address the limitations of existing approaches, we propose to utilize a point-of-view (POV) camera worn by an adult social partner to measure a child's eye contact behavior. This egocentric setting is ideally suited for social behavior measurement as the head-worn camera provides high quality visual data, e.g. consistent near-frontal views of faces, high resolution images of the eye regions, and less occlusion. Our approach capitalizes on the increased availability of POV platforms such as Pivothead and Snapchat Spectacles. In these platforms, a high definition outward-facing camera is integrated into a pair of glasses. For example, in the Pivothead, the camera is located at the bridge of the nose. Thus the camera is naturally aligned with the adult's eyes and when the child looks to the adult's eyes it will be captured as a look towards the camera (see Fig~\ref{fig:setup1}).

Detecting looks to the eyes in POV videos is still a challenging problem, however, due to the diverse and varied appearance of the human eye and the role of head pose in determining gaze direction. Specifically, the estimation of head pose is a critical factor in correctly interpreting the gaze direction based on analysis of the eye regions~\cite{land2009looking}. In our prior work on POV-based eye contact detection~\cite{ye2015detecting}, we addressed this issue by developing a set of machine learning-based eye contact detectors which were specialized for different configurations of head pose. Thus each detector was tuned for a particular range of child head poses. A strong disadvantage of this approach is the need to discretize the head pose and train independent detection models. In this paper, we remove this limitation through a novel deep learning solution in which head pose is jointly estimated along with eye contact detection.

This paper makes the following contributions:
\begin{enumerate}
\item We present a novel deep learning architecture for eye contact detection in POV video that outperforms previous approaches~\cite{ye2015detecting,smith2013gaze} by a large margin: 23.8\% gain over~\cite{ye2015detecting} and 50\% gain over~\cite{smith2013gaze} in $F_{1}$ score; 38.6\% gain over~\cite{ye2015detecting} and 64.5\% gain over~\cite{smith2013gaze} in area under the precision-recall curve
\item We conducted a systematic evaluation of our method on the largest corpus of naturalistic child social interactions available to-date, consisting of 22 hours of video from 100 children and encompassing four different interaction contexts and subjects with and without ASD
\item We will make our model for eye contact detection and our code for training and testing freely-available to the research community\footnote{See \url{http://cbi.gatech.edu/eyecontact/} for details.}
\end{enumerate}

\section{Related Work}
There are three categories of relevant prior work. First, we compare to the small number of previous works that addressed eye contact detection. Second, we describe recent works on large scale eye tracking that also make use of deep models for appearance analysis of eye regions. Third, we briefly review the use of classical eye tracking methods in autism.

\subsection{Eye Contact Detection}
Three prior works addressed the direct estimation eye contact from video and constitute the closest related work~\cite{ye2015detecting,shell2004ecsglasses,smith2013gaze}. The most relevant is our previous paper~\cite{ye2015detecting} which introduced the POV camera paradigm for eye contact detection. In comparison, this paper introduces a novel detection architecture based on deep learning which dramatically improves the detection performance. We also present the first experimental results for children with ASD, demonstrating that diagnostic status does not have a significant impact on detection performance. In addition, we present the first thorough experimental evaluation of our approach on 100 children across four different interaction contexts. Note that an abbreviated version of our prior POV work also appeared in~\cite{ye2012detecting}.

Shell et. al.~\cite{shell2004ecsglasses} developed an approach to eye contact detection based upon classical gaze tracking methods, using IR diodes on a pair of glasses to create glints on the eyes of the social partner. The reliance on special IR illumination greatly limits the usefulness of the method for naturalistic interaction. In their work on gaze locking~\cite{smith2013gaze}, Smith et. al. addressed a different application of eye contact, namely the use of gaze to an embedded camera as a user-interface technology in an internet-of-things context. Their approach predated the wide-spread use of deep models and their dataset consisted of subjects in a chinrest, with the consequence that they could not present results for naturalistic interactions. 

\subsection{Appearance-Based Gaze Estimation and Existing Datasets}
Traditional approaches to gaze estimation utilize active IR illumination to both create glints on the surface of the eye and reliably segment the pupil opening using a variety of dark and light pupil methods~\cite{hansen2010eye}. Recently, a number of investigators have explored alternative approaches to gaze estimation using appearance-based methods that analyze the eye region in conventional RGB images and avoid the use of structured illumination~\cite{bulling2017arxiv,krafka2016eye,sugano2014learning,zhang2015appearance}. We share with these methods the observation that the analysis of the eye region pattern in combination with head pose is a viable alternative to conventional gaze tracking technology. However, there are two main differences between our approach and these others. The first is that these prior works address the traditional eye tracking goal of determining the user's point of gaze on a display surface, as motivated by the widespread availability of user-facing cameras in tablets and laptop screens. These methods cannot be applied directly in our context of naturalistic face-to-face social interactions. Second, previously published methods~\cite{krafka2016eye,sugano2014learning,zhang2015appearance} utilize separately detected keypoints or eye region information as an additional input at testing time, whereas we present a full-face analysis approach which does not require any auxillary information at testing. The recent unpublished work~\cite{bulling2017arxiv} also describes a full-face analysis approach, but uses a different learning architecture from ours.

Like~\cite{bulling2017arxiv,krafka2016eye,zhang2015appearance}, we develop a deep CNN architecture to learn discriminative facial feature maps that encode gaze information. Unlike these works, we utilize the two stream architecture depicted in Fig~\ref{fig:picnn}, which is trained to output a head pose estimate along with the eye contact prediction. In contrast,~\cite{bulling2017arxiv} uses spatial weighting to combine CNN features, while~\cite{krafka2016eye} and~\cite{zhang2015appearance} use a separate estimation pipeline to compute facial landmarks and facial regions which are then combined with the deep CNN feature analysis. Our approach works directly with detected full-face regions and does not require externally-provided head pose estimates or landmark estimation, simplifying our analysis method significantly.

\begin{figure}[ht]
    \begin{subfigure}[t]{0.335\textwidth}
        \includegraphics[width=1\textwidth]{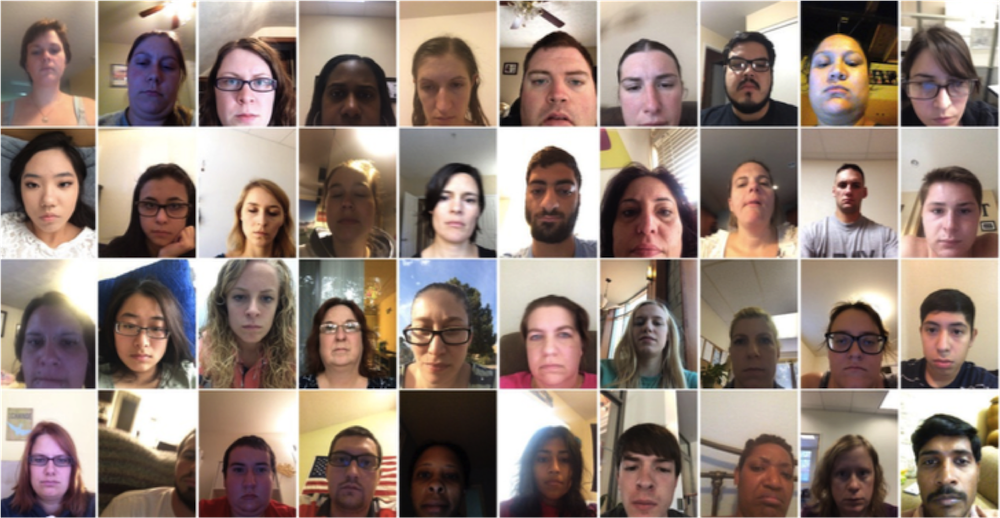}
        \caption{GazeCapture~\cite{krafka2016eye}}
    \end{subfigure}
    \begin{subfigure}[t]{0.245\textwidth}
        \includegraphics[width=1\textwidth]{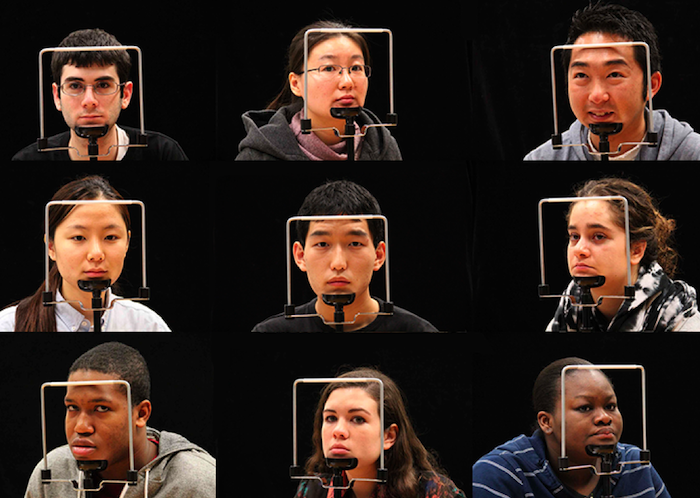}
        \caption{Columbia Gaze~\cite{smith2013gaze}}
    \end{subfigure}
    \begin{subfigure}[t]{0.41\textwidth}
        \includegraphics[width=1\textwidth]{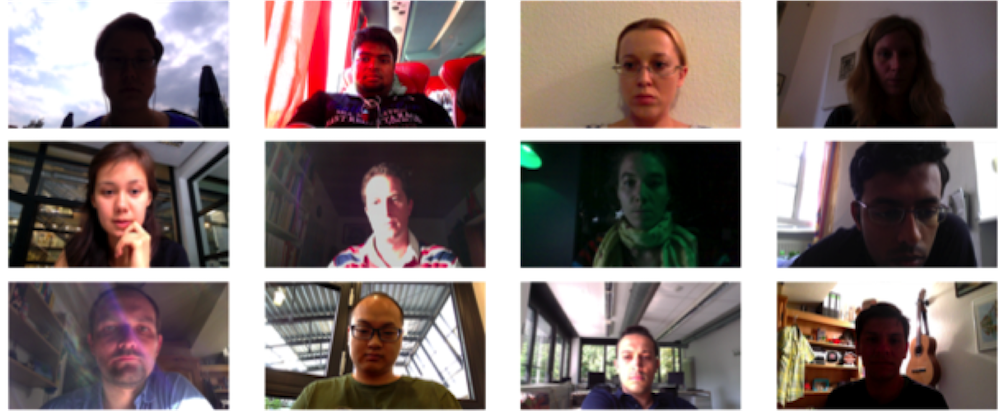}
        \caption{MPIIGaze~\cite{zhang2015appearance}}
    \end{subfigure}
  \caption{\textbf{Examples of publicly available gaze datasets.} As these datasets were collected from interactions with screens, they are not suitable for building a model that describes face-to-face social gaze behaviors.} \label{fig:3gazedataset}
\end{figure}

There are several publicly-available gaze datasets available to the research community. However, as these datasets were collected from interactions with screens, they are not suitable for building a model that describes face-to-face social gaze behaviors. Example images for the major public datasets are illustrated in Fig~\ref{fig:3gazedataset}. As these images demonstrate, the facial expressions and poses obtained from adult subjects interacting with screens are much less diverse than the variations in children's appearance and pose that we encounter in our setting (see Fig~\ref{fig:ecgallery} for examples). We will release our model trained with 22 hours of POV video consisting of social interactions with 100 child subjects. A subset of our training data is available as part of the MMDB dataset~\cite{rehg2013decoding}.\footnote{See \url{http://www.cbi.gatech.edu/mmdb/} for MMDB access instructions.}

\subsection{Gaze Behavior and Autism}
A significant amount of prior work has used eye tracking to investigate differences in patterns of looking in individuals with autism, such as reduced looks to social stimuli~\cite{chita2016social}. For example, toddlers with ASD spend more time looking at geometric shapes than human biological motion~\cite{pierce2011preference}, children with ASD devote less attention to faces while watching videos of social interactions~\cite{hosozawa2012children}, and both adults~\cite{klin2002visual} and children~\cite{chawarska2009looking} with ASD show preferential fixations to the mouth than to the eyes when viewing social scenes. However, all of these studies have been conducted in a highly-controlled environment in which the subjects were passively viewing a monitor screen for a short period of time. At present, we still lack a comprehensive understanding of similarities and differences in gaze behavior associated with autism~\cite{guillon2014visual}. Our work on eye contact can potentially complement this existing literature by providing insight into patterns of looking within a naturalistic social context.

Head-mounted eye tracking systems provide an alternative to monitor-based studies of gaze behavior and have been used in a limited number of studies involving children with ASD~\cite{noris2012investigating,magrelli2013social}. Findings from this work exhibit concordance with monitor-based gaze studies, but have also identified novel gaze patterns~\cite{noris2012investigating}, suggesting that more research is needed to understand gaze behavior in ecologically valid settings. A basic problem with using any form of eye tracking to analyze face-to-face gaze is the need to identify the gaze target given the estimated gaze direction. In other words, wearable eye tracking gives the location of the point of regard in a POV image, but does not directly answer the question of what gaze target is present at that location. This difficulty substantially increases the complexity of a fully-automated behavior measurement system based on wearable eye tracking. These issues, along with the challenges of compliance in requiring children to wear special hardware~\cite{sasson2012eye}, have limited the broad-scale applicability of this approach.

We note that a final, classical approach to obtaining social gaze measurements, including measurements of eye contact, is to manually annotate videos recorded in the lab setting or even home videos~\cite{zwaigenbaum2013early}. While this method is completely non-invasive and naturalistic, it is extremely time consuming and is subject to human error. Our previous work has shown that video from POV cameras is an effective medium for human annotation of children's social gaze~\cite{edmunds2017brief}. We utilize such annotations to construct the training and testing sets for our experiments.
 
\section{Method}
In order to detect eye contact in POV video we must first identify all of the faces that are present and then analyze each face separately to make a determination of eye contact. We describe our novel method for eye contact detection based on full-face analysis in Sec~\ref{sec:eyecontact_models}. This section assumes that all relevant faces have been detected. We describe our approach to face detection in Sec~\ref{sec:facedetection_methods}. In Sec~\ref{sec:facefiltering}, we describe an approach to identifying the detected face that corresponds to the child social partner, in the case where more than one face is present in an image. All of these analysis steps make use of modern deep learning architectures, yielding high accuracies relative to the previous state-of-the-art. We test our entire face analysis pipeline and separately evaluate each component in Sec~\ref{sec:result}.

\subsection{Eye Contact Detection}\label{sec:eyecontact_models}
Here we describe our main approaches to eye contact detection. Our prior work was based on a pose-dependent detection model which is summarized in Sec~\ref{sec:peec} for completeness. We then describe our new method based on a deep CNN architecture in Sec~\ref{sec:picnn}.

\subsubsection{Pose-Dependent Egocentric Eye Contact (PEEC) Detector} \label{sec:peec}

\begin{figure}[t]
  \centering
    \includegraphics[width=0.9\textwidth]{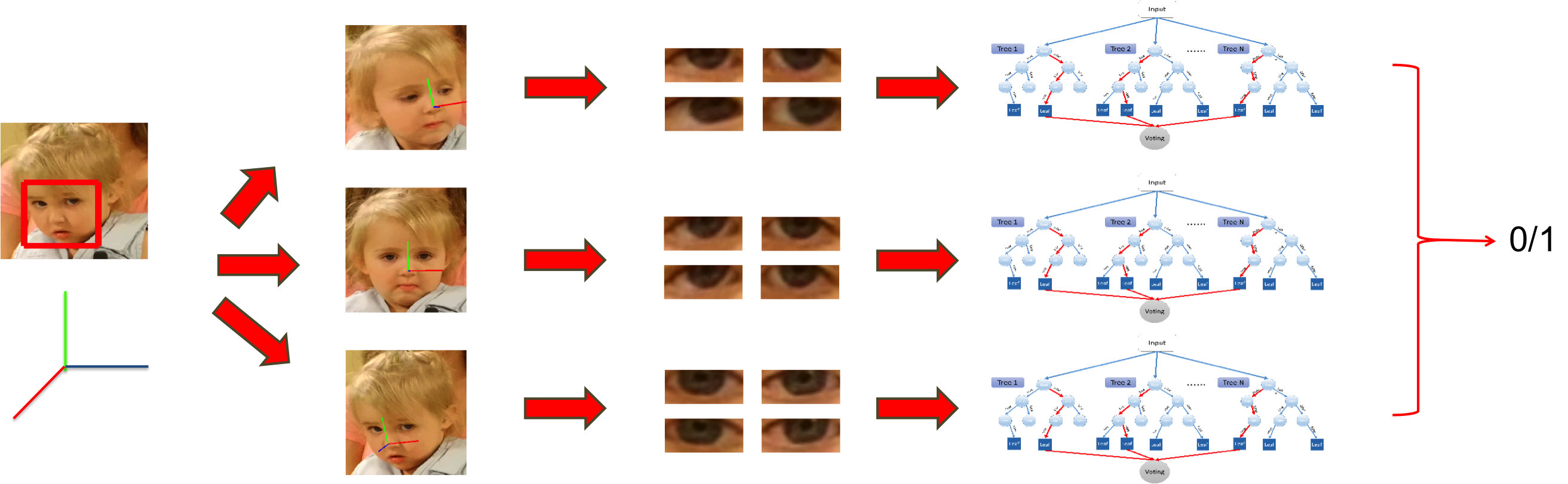}
  \caption{\textbf{Pose-dependent egocentric eye contact detector (PEEC).} From an input image, face is detected and head pose estimated. Image is assigned to one of three clusters based on head pose, eye features are extracted, and cluster-specific random forest classifier is used to predict the final output.}\label{fig:rf_pipeline}
\end{figure}

Fig~\ref{fig:rf_pipeline} illustrates the pipeline for the PEEC method we introduced in~\cite{ye2015detecting}. The input is an egocentric video frame and the output is a prediction of eye contact. Face detection is used to extract a bounding box containing the child's face. Given a bounding box, we identify facial landmark points and estimate the head pose using the IntraFace method from~\cite{de2015intraface}. IntraFace extends cascaded pose regression~\cite{dollar2010cascaded} with a Supervised Descent Method for tracking facial landmarks. The facial landmarks are then used to estimate the head pose with three degrees of freedom- yaw, pitch and roll. These preprocessing stages are illustrated in Fig~\ref{fig:prep}. Using the results of the facial landmark identification, we crop the eye regions and extract Histogram of Oriented Gradients (HOG) features~\cite{felzenszwalb2010object} from the cropped eye patches. Using the estimated head pose, the training data is divided into three groups and a binary classifier for eye contact is trained for each group. At testing time, all three classifiers are applied to each frame and the results are aggregated. We now describe each of these steps in more detail.

Head Pose Clustering - This is an important step because the appearance of the human eye during moments of eye contact varies with the head pose. The head pose of the child's face $h_t$ is estimated at frame $t$ by registering the InfraFace landmarks onto an average 3D face. We cluster head poses using a Gaussian mixture model and train a separate classifier for each pose cluster. Since rotation can be resolved by affine transformations, only the pitch and yaw are used for head pose clustering, with the number of clusters set to 3 for all the experiments.

Feature Extraction - 
Left and the right eye patches from each face are cropped using IntraFace facial landmarks and an affine transformation is used to align tilted eyes. We resize all of the images to a fixed size of 73 x 37. We extract HOG appearance features $A_t$ from both the left and right eyes at frame $t$ and concatenate them to create a feature vector. 

Eye-Contact Detection - We assign a label $y_t$ to each frame $t$, where $y_t$ is 1 if there is eye contact and 0 otherwise. Let the appearance features be denoted by $A_t$ and the head pose by $H_t$ for the frame $t$.
Then the conditional probability $P(y_t| H_t, A_t)$ is given by:

\begin{figure}[ht] 
  \centering
    \includegraphics[width=0.19\textwidth]{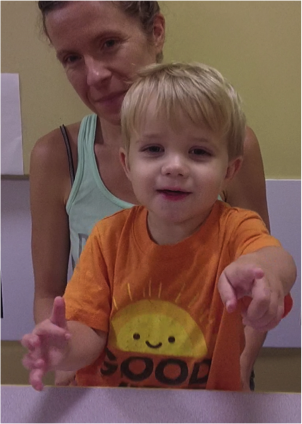}
    \includegraphics[width=0.19\textwidth]{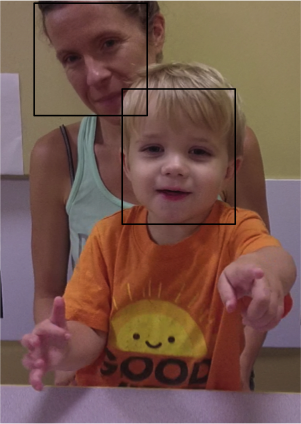}
    \includegraphics[width=0.19\textwidth]{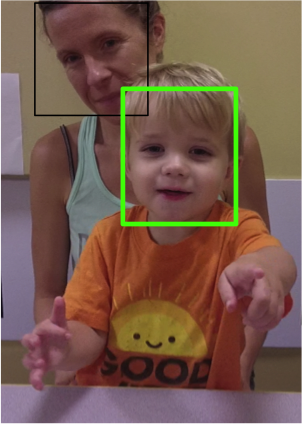}
    \includegraphics[width=0.19\textwidth]{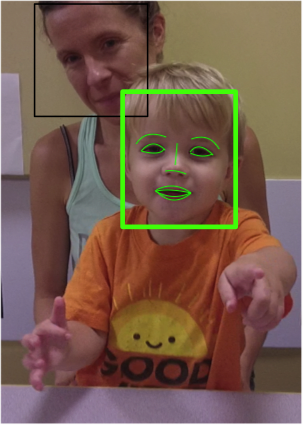}
    \includegraphics[width=0.19\textwidth]{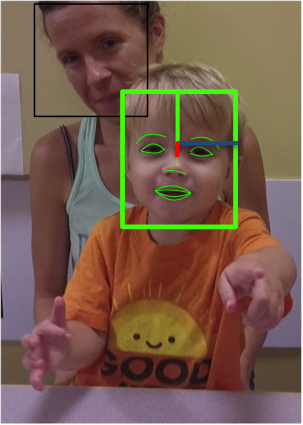}    
  \caption{\textbf{Preprocessing steps in PEEC prior to eye contact detection.} Left to right: 1) Input image; 2) All faces are detected; 3) Child's face is selected; 4) Facial landmarks are localized; and 5) Head pose is estimated.} \label{fig:prep}
\end{figure}

\[P(y_t| H_t, A_t) = \Sigma^{k}_{h_c=1}P(h_c|H_t)P(y_t|h_c,A_t),\]
where $h_c$ enumerates the head pose clusters in the Gaussian mixture model. Each cluster is considered as a sub-problem and the classifier outputs from each cluster are averaged together. We utilize a Random Forest classifier~\cite{breiman2001random} with 100 trees and 10 splits per node.

\subsubsection{Pose-implicit Convolutional Neural Networks (PiCNN) Detector} \label{sec:picnn}

There are two major problems with the PEEC approach from Sec~\ref{sec:peec}. First, due to the fact that the method requires head pose estimates (to assign each sample to one of the pose clusters) and eye localizations (to extract features), gaze estimation cannot be performed when landmark detection fails. Thus any failure in steps 4 and 5 of the preprocessing pipline of Fig~\ref{fig:prep} will result in a missed detection.
Facial landmark detection is a more difficult problem than face detection. It is challenging to detect the landmarks reliably when the face is occluded or foreshortened due to out-of-plane rotation, which happens quite frequently during dynamic social interactions with children (see Fig~\ref{fig:intrafail}). Our experiments reveal that, in our dataset, landmarks are successfully found in only 75.56\% of the total eye contact frames, whereas faces are detected 97.94\% of the time (Table~\ref{tab:intra_result}). The second problem with PEEC is its reliance on hand-designed HOG features and a random forest classifier, which are known to be inferior to modern deep neural network classifiers which support end-to-end feature learning.

We therefore propose a new classification architecture based on Convolutional Neural Networks (CNN)~\cite{krizhevsky2012imagenet} that addresses these two issues. Our new approach learns an image representation that \emph{jointly} predicts head pose and eye contact, instead of being dependent on precomputed head pose and facial landmarks. Since the representation is learned end-to-end (i.e., the input is raw pixels and the output is a binary prediction) there is the opportunity to learn features that are specific to the task. This approach is particularly promising in light of the large-scale social interaction dataset that we have assembled. Our proposed model is called \textit{``Pose-implicit CNN Detector''} or \textit{PiCNN}. 

Fig~\ref{fig:picnn} illustrates the PiCNN architecture. The input to the network is a rectangle of pixels corresponding to a face bounding box produced by a face detector (see Sec~\ref{sec:facedetection_methods}). The detected face patch is resized to $227\times227$. The network has five convolutional and pooling layers, similar to AlexNet~\cite{krizhevsky2012imagenet}, but with a smaller filter size (7 by 7) with strides of 2 in the first layer to capture the finer details in the face. Layers 6 to 8 are fully connected. The primary difference between our approach and AlexNet is the presence of two branches in the 7th and 8th fully-connected layers. The upper branch outputs a prediction of the three axes of head rotation (i.e., regression in yaw, pitch, roll) and the lower branch is tasked with the binary classification of eye contact. 
The weights and parameters are the same and are linked during training in the first 6 layers (5 convolutional and 1 fully connected) and branched in the last 2 layers to facilitate multi-task learning. 
Thus the prediction of eye contact can benefit from implicitly learning the variability in eye appearance resulting from head pose change. This model achieves the best performance, by a large margin, among all existing methods on our dataset, as presented in Sec~\ref{sec:eyecont_result}. 

In training the PiCNN model, we have ground truth eye contact labels for every frame in the training dataset. This allows us to backpropagate training error through the eye contact detection branch in every training batch.\footnote{See~\cite{Goodfellow-et-al-2016} for details on neural network training in general, and Sec~\ref{sec:eyecont_result} for the details in our approach.} However, we do not have frame-level human annotations of head pose for any frames. We solve this problem by using the IntraFace system as a source of head pose training data. In frames for which head pose estimates from IntraFace are available, we additionally backpropagate training error through the head pose branch. Note that this strategy has the benefit that the network is forced to learn representations for predicting eye contact even when a head pose reference is not available, making it potentially more robust than a sequential approach in which head pose must be computed before detection can be performed. At the same time, our method can take advantage of sparse head pose annotations where they are available and improve the detection performance. During testing time, we apply the PiCNN model in feedforward mode and do not utilize any separate head pose estimates. Now that we have presented our two approaches to eye contact detection, PEEC and PiCNN, we briefly describe the related methods which are used as baselines in our experiments in Sec~\ref{sec:result}.

\subsubsection{Modified AlexNet} \label{sec:alexnet}
\begin{figure}[t]
  \centering
    \includegraphics[width=0.9\textwidth]{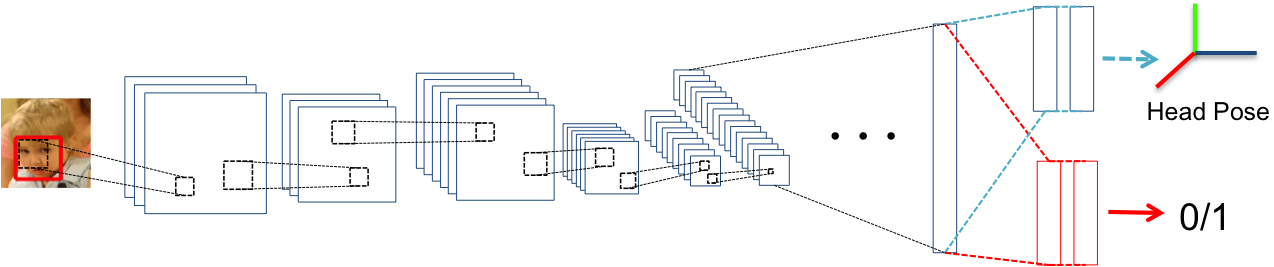}
  \caption{\textbf{Pose-implicit Convolutional Neural Networks (PiCNN) architecture for eye contact detection.} We utilize a two stream network structure with 8 layers modeled after AlexNet. One output branch predicts head pose and the other predicts eye contact. The inputs to the network are face bounding boxes. A representative face bounding box is illustrated with a red outline in the image at the left.}
\label{fig:picnn}
\end{figure}

One important performance baseline for our method is the architecture in Fig~\ref{fig:picnn} with the head pose prediction branch removed. Modulo some small differences in the convolutional layers, this is equal to the standard AlexNet implementation~\cite{krizhevsky2012imagenet}. This approach requires the model to learn features which encode head pose cues without having access to an explicit training signal. Our performance gains over this model demonstrate the utility of our two branch approach and the benefit of providing an explicit reference for head pose during training.

\subsubsection{Gaze Locking}
We have implemented the Gaze Locking method from~\cite{smith2013gaze} to compare its performance against our method. Using the facial landmark detection results, we detect the location of the eyes in each frame. Eye tilt is corrected using affine transformations and each eye is cropped to a size of 37 x 73. We then concatenate the intensity pixels from the left and the right eye to form a high dimensional feature vector. This is then projected onto a low dimensional space by performing Principal Component Analysis, (PCA)~\cite{turk1991face}, followed by Multiple Discriminant Analysis, (MDA)~\cite{duda2012pattern}. The resultant feature vector is then fed into a support vector machine~\cite{chang2011libsvm}, which is trained offline on the training set. We choose to reduce our high dimensional feature vector into 200 dimensions using PCA. We find that compressing this further with MDA into a 6 dimensional subspace gives good results. Our support vector machine is used with default parameters as described in the original paper. The SVM performs a binary classification on each frame.

\subsection{Face Detection} ~\label{sec:facedetection_methods}
All of the eye contact detection methods described above require a detected face bounding box as input. Here we describe the approach to face detection used in our experiments. There exists a substantial literature on face detection, with approaches such as~\cite{viola2004robust} seeing widespread use prior to the advent of deep learning. In the same way that deep CNNs have come to dominate the fields of object recognition and object detection, face detection has similarly turned to CNNs, yielding impressive results in benchmark datasets such as \cite{jain2010fddb} and \cite{yang2016wider}. Recently, complex pipelines have been built for this task to deal with issues such as distant faces and low quality images~\cite{yang2015facial,yang2016wider,zhang2016joint}. Since we do not encounter these problems in our dataset, we adopt a simpler approach.

The general purpose object detection method Faster R-CNN~\cite{ren2015faster} achieved impressive performance in object detection challenges such as Microsoft COCO~\cite{lin2014microsoft} and PASCAL VOC~\cite{everingham2010pascal}. Faster R-CNN generates object proposals from convolutional feature maps of the image using a trainable Region Proposal Network. These proposals are then evaluated by the Fast R-CNN detection network which outputs high confidence object detections. Jiang et al.~\cite{jiang2016face} demonstrated that Faster R-CNN can be trained to obtain a highly-accurate face detector. Similarly to~\cite{jiang2016face}, we train Faster R-CNN on the WIDER FACE training dataset. We analyze our entire dataset using this detector and output a list of face bounding boxes for every frame.

\subsection{Child Face Selection} \label{sec:facefiltering}
In general, face detectors will identify all faces in the environment, including a parent's face (see Fig~\ref{fig:prep}), a sibling's face, and occasional face-like spurious patches, along with the child's face. Since we are only interested in the child's social behavior, we need to select the child's face and reject the other face patches. To classify individual faces we adopt an online appearance learning approach, where in the beginning a pre-defined number of face classes are initialized (ranging from one to four in our experiments), a simple classifier is learned, in the next frame each detected face is added to one of the classes, and then the classifier is updated and the process is repeated. For facial appearance feature extraction, we use the VGG-face model~\cite{parkhi2015deep}, a pre-trained deep network for face recognition which is well-matched to our application. For online classification, we use logistic regression with sequential updates. The combination of deep network features with on-line classifier learning successfully identifies the child's face at very high precision and recall (both greater than 90\%, details in Table~\ref{tab:vatic_result} and~\ref{tab:intra_result}).

\section{Datasets}
\label{sec:datasets}

\begin{figure}[t]
  \centering
    \includegraphics[width=0.7\textwidth]{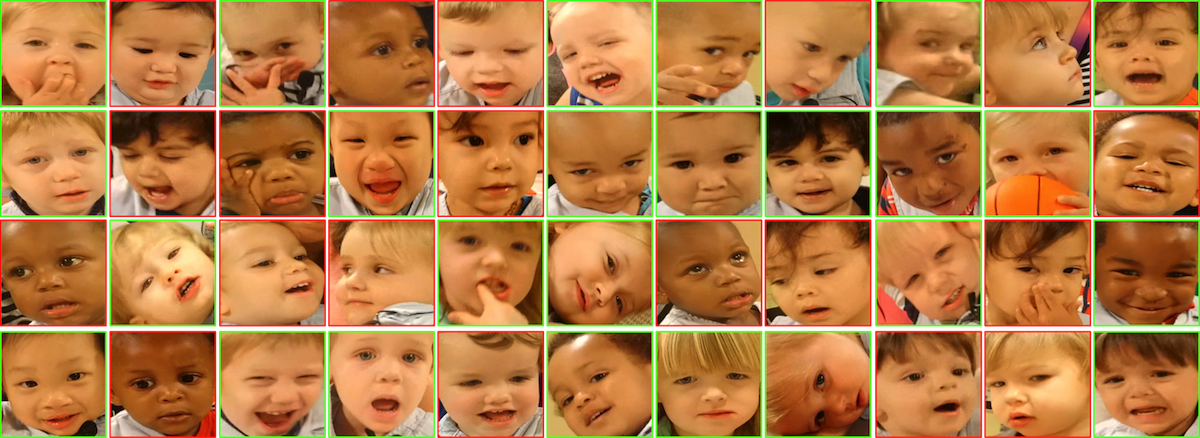}
  \caption{\textbf{Sample child faces in our dataset.} Green box indicates ground truth eye contact, red indicates no eye contact. Notice the diversity and richness in the children's facial expressions and head pose during natural social interactions.} \label{fig:ecgallery}
\end{figure}

The data utilized in our experiments comes from four separate studies, conducted at one or more of the following three sites: the Georgia Tech Child Study Lab in Atlanta, GA (GT); the Center for Autism and the Developing Brain in White Plains, NY (CADB); and the Marcus Autism Center in Atlanta, GA (MAC). Descriptive information for the subset of participants whose data was included in the current analysis and details of the data collection protocol for each study is detailed below, separately for each dataset, and in aggregate in Table~\ref{tab:dataset_sum}. Sample images from our dataset are illustrated in Fig~\ref{fig:ecgallery}.

Generally, all four data collection protocols involved a semi-structured play interaction between an adult and a child, who sat across a small table from each other. The specific protocols chosen were selected based on their prior use in research on social attention and communication in typically developing children and children with autism, and because they have been shown to reliably elicit eye contact from these groups. In all four studies, the examiner interacting with the child wore a pair of commercially-available glasses - Pivothead Kudu - which have an outward-facing camera embedded in the bridge over the nose. By virtue of its placement, this camera reliably captures a close-up image of the child's face as the examiner interacts with the child. The lenses were removed from the glasses to provide the child an unobstructed view of the examiner's eyes. Our prior research indicates that the presence of the glasses does not affect the gaze behavior of children~\cite{edmunds2017brief}.

Research assistants annotated the pivothead videos from each dataset using one of two video-annotation software tools: ELAN\footnote{http://tla.mpi.nl/tools/tla- tools/elan/} and INTERACT Mangold, 2017.\footnote{https://www.mangold-international.com/en/products/software/behavior-research-with-mangold-interact} Ground truth coding involved flagging the frame-level onset and offset of each instance of the child making eye contact with the examiner, as captured by looks into the camera. Kappas for frame-level agreement between pairwise comparisons of the 6 coders ranged from .89 to .94.

All studies were approved by the Institutional Review Boards of the respective institutions. Caregivers of the children participating in the studies provided written consent for their child's participation, video recording, and data sharing.

\subsection{Dataset 1}
Twenty-eight children with a diagnosis of Autism Spectrum Disorder (ASD) (3 females) between the ages of 5 years and 13.7 years (mean age = 7.2 years) were recruited through the Center for Autism and the Developing Brain (CADB) in White Plains, NY. The sample was 67.9\% Caucasian, 17.9\% mixed ethnicity, 3.6\% Hispanic, and 3.6\% Asian (2 missing). All participants completed the play-based assessment during a single visit; a subset (n = 15) participated in a second assessment, eight weeks from their initial visit.
Diagnosis of ASD was confirmed prior to participation by a licensed clinical psychologist at CADB. A best estimate diagnosis was based upon information collected from the Autism Diagnostic Observation Schedule (ADOS)~\cite{lord2012autism} and the Autism Diagnostic Interview-Revised (ADI-R)~\cite{rutter2003autism}.
All participants were recorded in a modified version of the Brief Observation of Social Communication Change (\textbf{v-BOSCC})~\cite{verbalboscc}, a 12-minute examiner-child interaction that consisted of two 5-minute play segments with standardized sets of toys, separated by a 2-minute conversation segment. The examiner and participant were seated across from each other (face-to-face) at a small table. During the play, the child was given the option to choose a single toy from a box that contained a standardized set of toys. The set of toys available in the box differed between the first play segment and the second play segment. Within each play segment, the child was free to chose a new toy if they no longer wanted to play with the toy originally chosen. The child was free to play with each toy as he or she saw fit; the examiner joined the child's play but did not guide it, and maintained an amount and level of language commensurate with the child's. The transition to the conversation segment was signaled by the examiner stating that it was time to clean up and then introducing an open-ended conversation topic (e.g., ``I went to the park this weekend''). During the conversation, no toys were present on the table.
\subsection{Dataset 2}
Two groups of participants were recruited for this study: a sample of typically developing children between the ages of 18 and 36 months recruited at Georgia Tech, and a sample of 3-6 year old children with a diagnosis of ASD recruited at CADB. The sample included in the current analysis consisted of 19 children with ASD (6 females) and 16 typically developing children (7 females). The ASD sample ranged in age from 23 to 60 months (mean = 45 months), and was 63\% Caucasian, 15.7\% Asian, 10.5\% Hispanic, and 5.2\% mixed ethnicity (1 missing). The TD sample ranged in age from 20 to 36 months (mean = 28.3 months), and was 81.3\% Caucasian, 12.5\% mixed ethnicity, and 6.3\% African American.

Children participated in two assessments, described in more detail below: the nonverbal version of the Brief Observation of Social Communication Change (\textbf{nv-BOSCC}; 13 ASD, 13 TD) and the Early Social Communication Scales (\textbf{ESCS}; 13 ASD, 14 TD). 10 ASD and 11 TD children contributed data to both assessments.

The ESCS~\cite{mundy2003early} is a 15-25 minute structured assessment that uses standardized toys, examiner-initiated prompts (e.g., points to pictures in a book and posters on the wall) and contextual presses (e.g., wind-up toys activated out of reach of the child) to elicit gaze shifts relevant to different communication functions (sharing attention, requesting, maintaining social interaction). The examiner presents the toys to the child, one a a time, first activating the toy (e.g., blowing up balloon and then slowly letting the air out to make a squeaking noise; pressing on a trapeze toy to make a monkey swing around; winding up small toys that move across the table) and then handing the toy to the child before retrieving the toy and activating it two more times. This assessment has been shown to reliably elicit shifts of attention from objects to the examiner's eyes in typically developing children and children with autism~\cite{sigman1986social,rozga2011behavioral}.

As described above for dataset 1, the BOSCC is a naturalistic examiner-child play interaction that involves two play segments involving toys, separated by a segment of non-object-mediated social interaction. Due to the younger age of the children participating in this study, the nonverbal version of the BOSCC was used~\cite{grzadzinski2016measuring}. This version follows the same procedures as described above in dataset 1, but utilizes more age-appropriate materials and activities. Hence, the boxes include toys selected based on their appropriateness for younger children, and the length of play per box is reduced to 4 minutes. The conversation segment is replaced with a 2-minute snack in which the examiner offers the child a choice of crackers and cookies and then engages the child socially by commenting on the activity (e.g., ``Goldfish crackers, yum yum!''). During the snack, no toys are present on the table. The transition from the first toy play segment to the snack segment is signaled by the examiner stating it was time to clean up for snack, and the transition from the snack to the second toy play segment is signaled by the examiner stating ``Let's play with some new toys.''

\begin{table}[t]
\centering
\resizebox{\columnwidth}{!}{%
\begin{tabular}{rr|rr|rr|rr|rr|rr}
\multicolumn{2}{c|}{age (in months)} & \multicolumn{2}{c|}{gender} & \multicolumn{2}{c|}{diagnosis} & \multicolumn{2}{c|}{ethnicity} & \multicolumn{2}{c|}{protocol} & \multicolumn{2}{c}{annotation} \\ \hline
                   & \# subjects    &            & \# subjects    &            & \# subjects       &                  & \# subjects &              & \# sessions    & \# frames          & minutes       \\
less than 20       & 8              & male       & 74             & TD         & 50                & Caucasian        & 60          & ESCS         & 30             & 2,364,773       & 1,314         \\
20 $\sim$ 29       & 33             & female     & 26             & ASD        & 50                & mixed            & 16          & R-ABC        & 34             &                 &               \\
30 $\sim$ 39       & 15             &            &                &            &                   & African American & 10          & v-BOSCC      & 43             &                 &               \\
40 $\sim$ 49       & 7              &            &                &            &                   & Asian            & 6           & nv-BOSCC     & 26             &                 &               \\
50 $\sim$ 59       & 4              &            &                &            &                   & Hispanic         & 4           & Marcus       & 23             &                 &               \\
60 $\sim$ 69       & 14             &            &                &            &                   & unknown          & 4           &              &                &                 &               \\
70 $\sim$ 79       & 5              &            &                &            &                   &                  &             &              &                &                 &               \\
80 $\sim$ 89       & 5              &            &                &            &                   &                  &             &              &                &                 &               \\
greater than 90    & 9              &            &                &            &                   &                  &             &              &                &                 &              
\end{tabular}%
}
\caption{Dataset summary.\label{tab:dataset_sum}}
\end{table}

\subsection{Dataset 3}
Typically developing children between 15 and 36 months of age were recruited from the community at large in metropolitan Atlanta via fliers posted at daycare centers and a mailing to parents of children in the appropriate age range using a mailing list purchased from Experian. The sample included in the current analysis consists of 34 children (9 females) who ranged from 16 to 34 months (mean = 23.9 months). The sample was 53\% Caucasian, 23.5\% African American, 17.6\% mixed ethnicity, and 5.9\% Asian. Twelve children in the current analysis were also included in the analyses we reported in~\cite{ye2015detecting}.

Children participated in the Rapid-ABC assessment (\textbf{R-ABC})~\cite{ousley2013beyond}, a 2-4 minute structured interaction led by the examiner that consists of five activities: (1) examiner greets the child; (2) examiner presents a ball to the child and then initiates a turn-taking game by rolling the ball back and forth across the table; (3) examiner presents a book to the child and then reads it out loud while encouraging the child to turn the pages and to point to pictures; (4) examiner surprises the child by placing the book on her head as if it were a hat and waits for the child's reaction; and (5) the examiner initiates a gentle tickle game with the child. More details about this assessment can be found in Rehg et al~\cite{rehg2013decoding}.

\subsection{Dataset 4}
As part of a pilot study to trial the use of the wearable camera in a clinical setting, data was collected from three boys with a diagnosis of autism between the ages of 3 and 6 years. One boy was Caucasian, one was African-American, and the ethnicity of the third is unknown. Participants were recruited from the Language \& Learning Clinic (LLC) at the Marcus Autism Center, where all were receiving intervention services at the time. Data collection involved sessions intended to resemble treatment sessions at the clinic, including sessions where the therapist engaged the child in toy play while placing no demands on the child to respond (Pairing, 12 sessions), sessions in which the therapist placed demands on the child, (Demands, 5 sessions), and sessions where the therapist periodically withheld toys from the child in an effort to get the child to request the toy (Mands; 6 sessions). We denote these plays as~\textbf{Marcus} in this paper. All but two sessions were 10 minutes in length; the remaining two were 3 and 9 minutes in length, respectively.

\subsection{Dataset Summary}
In summary, when all of the four datasets are considered together, there are 100 unique children in 156 different play sessions (some participants completed two sessions). Half of the children were diagnosed with ASD and 74\% of the participants were boys. The total number annotated frames exceeds 2 million, corresponding to approximately 22 hours of video. The complete statistics for this dataset are given in Table~\ref{tab:dataset_sum}.

\section{Experiments and Results} \label{sec:result}
In this section, we provide experimental evaluations for each of the components of our method.

\subsection{Child Face Detection Results} \label{sec:face_detction_result}
In order to evaluate the accuracy of face detection, we performed additional annotations in a sample of frames. We annotated child faces with bounding boxes for a total of 5 minutes of video containing a single child participant, and 5 minutes of video containing a child and a caregiver together, corresponding to approximately 18,000 frames in total. Using these annotations as ground truth, we compared the accuracy of the Faster R-CNN face detector from Sec~\ref{sec:facedetection_methods} to the OMRON OKAO face detector~\cite{omron} which was used in our previous experiments in~\cite{ye2015detecting}. We obtained detections for every frame using both methods, and filtered spurious false positives and adult faces for both methods using the online selection method from Sec~\ref{sec:facefiltering}. The precision-recall results at Intersection over Union (IoU) thresholds of 0.5 and 0.75 are given in Table~\ref{tab:vatic_result}. IoU score is defined as overlapping ratio between the predicted and ground truth bounding boxes. See Sec. 4.2 in~\cite{everingham2010pascal} for a description of IoU score and its use in evaluating object detectors.
The results show a high recall discrepancy between the two methods, with Faster R-CNN delivering substantially greater recall at a comparable precision for IoU 0.5. High recall is a critical property for our application, in order to not miss brief moments of eye contact.

\begin{table}[t]
\centering 
\begin{tabular}{ccc|cc} 
IoU threshold                     & \multicolumn{2}{c|}{0.5} & \multicolumn{2}{c}{0.75} \\ \hline
\multicolumn{1}{c}{}             & Recall    & Precision    & Recall     & Precision    \\
\multicolumn{1}{c}{OMRON}        & 0.437     & 0.941        & 0.015      & 0.212        \\
\multicolumn{1}{c}{Faster R-CNN} & \textbf{0.947}     & \textbf{0.963}        & \textbf{0.828}      & \textbf{0.870}       
\end{tabular}
\caption{Precision and recall comparison at two IoU thresholds between the OMRON OKAO face detector and Faster-RCNN trained on the WIDER FACE dataset. Testing set consists of approximately 18k child face examples.\label{tab:vatic_result}}
\end{table}

\subsection{Landmark Detection and Pose Estimation Results} \label{sec:intraresults}
\begin{figure}[t]
  \centering
    \includegraphics[width=0.19\textwidth]{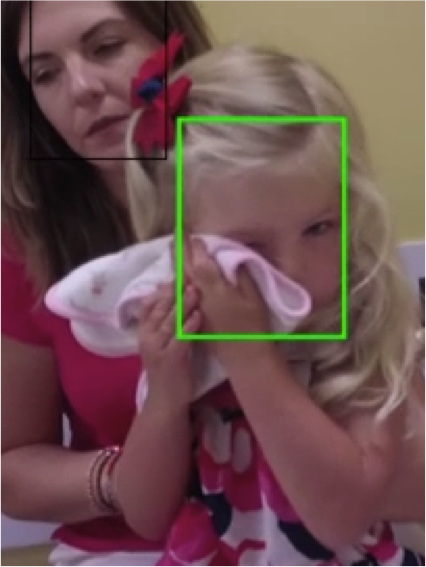}
    \includegraphics[width=0.19\textwidth]{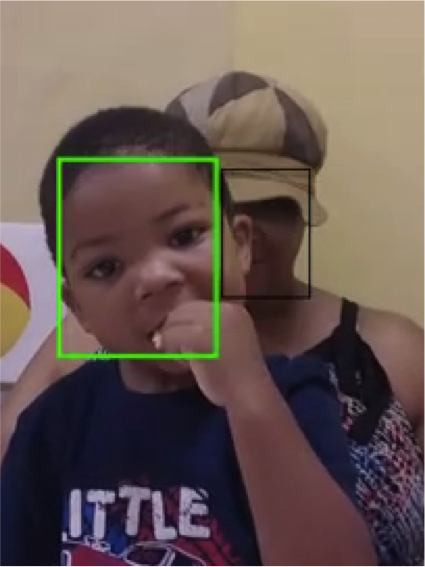}
    \includegraphics[width=0.19\textwidth]{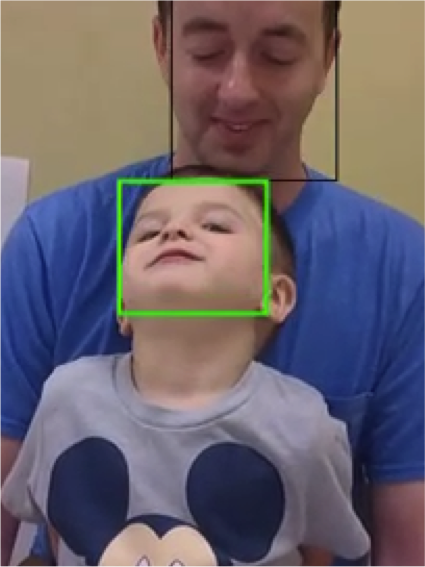}
    \includegraphics[width=0.19\textwidth]{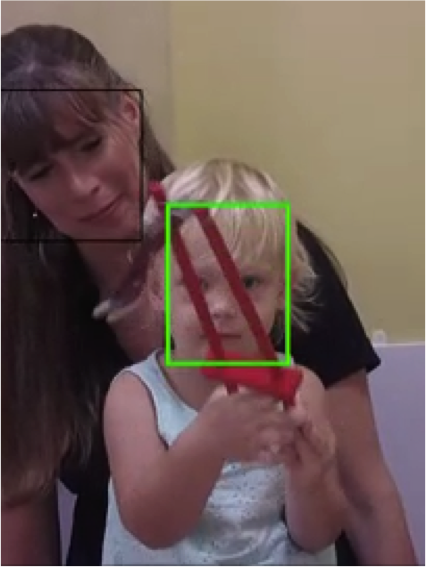}
    \includegraphics[width=0.19\textwidth]{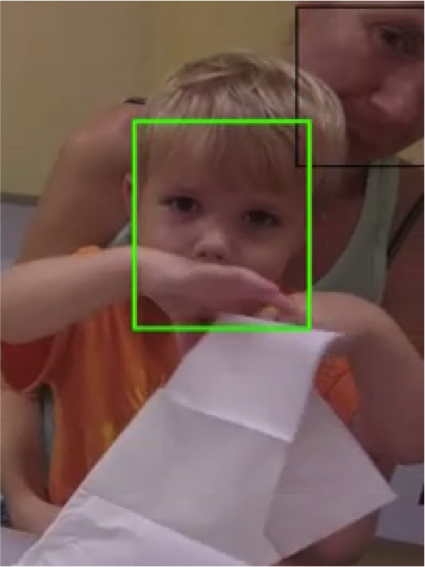}
  \caption{Landmark detection failures due to occlusions or extreme head pose.} \label{fig:intrafail}
\end{figure}

To evaluate the frequency of successful landmark detection (and by extension, head pose estimation) in our dataset, we calculated the percentage of frames labeled as eye contact for which landmarks were detected, shown in Table~\ref{tab:intra_result}. Note that this is not a statement about the accuracy of the landmarks, only about their availability. This is meaningful because state-of-the-art methods such as IntraFace will only output landmarks if they are of sufficiently high quality. The percentage of frames with missing landmarks is the portion of the error rate for eye contact detection which is attributable to landmark detection failure. As can be seen in Table~\ref{tab:intra_result}, landmark detection is successful in only 75.56\% of frames, in comparison to 97.94\% for face detection. This suggests that around a quarter of the errors are due to landmark detection failure, and a negligible amount are due to face detection failure. These findings explain the low recall rate of the two eye contact methods (PEEC and GazeLocking) that rely on landmark detection. Fig~\ref{fig:intrafail} shows some examples of situations of occlusion and extreme rotation that result in landmark detection failures.

\subsection{Eye Contact Detection Results} \label{sec:eyecont_result}

We now describe our experimental evaluation of four eye contact detection methods, PiCNN, PEEC, Modified AlexNet, and GazeLocking, using the entire dataset from Sec~\ref{sec:datasets}. 

\subsubsection{Experiment Design}
We divide the total dataset summarized in Table~\ref{tab:dataset_sum} into 5 disjoint train/test splits for 5-fold cross validation, where each split has 80\% of training and 20\% of testing sessions and subjects included in the training set are not present in the testing set. This ensures that testing is always done on unseen participants.
Each training set split is sampled uniformly across combinations of diagnosis (TD/ASD) and play protocols. For example, we sample 80\% for a training set from the TD ESCS group, from the ASD ESCS group, from the TD nv-BOSCC group, from the ASD nv-BOSCC group, and so on, such that different conditions are fairly represented across the five splits. We use the same five folds to train, test and compare the four eye contact detection models described in Sec~\ref{sec:eyecontact_models} -- our PiCNN model, PEEC~\cite{ye2015detecting}, Modified AlexNet~\cite{krizhevsky2012imagenet}, and Gaze Locking~\cite{smith2013gaze}. With this cross validation approach, we are able to obtain testing results on all of our datasets, and we report the overall performance averaged over all sessions as well as within different populations. Table~\ref{tab:final_scores} summarizes the results.

On average, the training set in each split initially had 145k positive (eye contact) frames and 1,746k negative (no eye contact) frames, which is highly imbalanced. To overcome the data skewness issue, we resampled the training sets with horizontal flip, slight rotation and color jittering with positive set oversampling and negative set subsampling to make it more balanced at positive:negative = 561k:842k = 4:6 ratio. 

\begin{table}[t]
\centering 
    \begin{tabular}{l|rr|rrrrr||r}
    ~                       	          & ASD & TD 			& ESCS 		& v-BOSCC & nv-BOSCC 	& R-ABC	& Marcus 		& All \\ \hline
    total frames                     & 118,464   & 62,220  	& 29,305		&  74,867         	& 38,891      	& 19,718	& 17,903      & 180,684   \\ 
    face detected (\%), mean & 98.21   & 97.52  		& 97.45		& 98.90  & 95.34  			& 98.30	& 99.32  		& \textbf{97.94}  \\
    face detected (\%), std     & 0.03     & 0.03    		& 0.03		&  0.02   &  0.05   			& 0.02	&  0.01   		&  0.03   \\ 
    landmark detected (\%), mean & 73.65   & 70.93  	& 81.23		& 80.06  & 80.23  			& 59.99	& 55.77  		& \textbf{75.56}  \\
    landmark detected (\%), std     & 0.19     & 0.22    	& 0.14		&  0.13   &  0.16   			& 0.24	&  0.21   		&  0.21   \\
    \end{tabular}
    \caption{Total number of frames within the moments that are marked as eye contact by human annotators, and the success rate (mean and standard deviation) of face detection and landmark detection, grouped by different conditions.\label{tab:intra_result}}
\end{table}

For the two deep models (PiCNN and AlexNet), we used the Caffe deep learning framework~\cite{jia2014caffe} on an Nvidia Titan X 12GB GPU, with batch size 128, learning rate 0.005 until the 100,000th iteration and 0.0005 until the final 200,000th iteration. For the other two models (PEEC and Gaze Locking), we used MATLAB on an Intel Core i7-5930K CPU.

Note that our analysis does not include direct performance comparisons to other appearance-based gaze tracking methods such as~\cite{krafka2016eye,zhang2015appearance}. The primary reason is that these methods are designed for a different task, accurately mapping the participant's point of gaze on a mobile screen. These methods can identify \emph{where} a participant is looking, but they only know \emph{what} the user is looking at if they have access to ROI masks for the gaze stimulus. In contrast, our method automatically identifies \emph{what} the participant is looking at, but only for the specific gaze event of eye contact. Moreover, the domain of screen viewing is quite different from naturalistic social interactions. This can be seen by comparing Figs~\ref{fig:3gazedataset} and~\ref{fig:ecgallery}. In particular, \cite{krafka2016eye,zhang2015appearance} rely on landmark localization~\cite{baltruvsaitis2014continuous} to extract the eye region from the face. Based on our results from Sec~\ref{sec:intraresults}, these methods will take an immediate 22.4\% miss in recall (see Table~\ref{tab:intra_result}) due to the difficulty of detecting landmarks under challenging conditions. These methods are therefore unlikely to be competitive with PiCNN for our task.

\subsubsection{Experimental Results}

\begin{figure}[!tbp]
  \centering
  \begin{minipage}[b]{0.49\textwidth}
    \includegraphics[width=\textwidth]{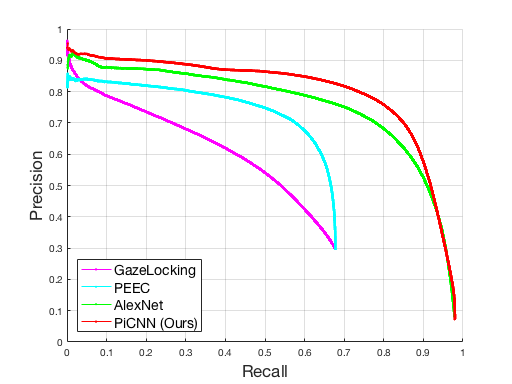}
    \caption{Average Precision-Recall curve.}
  \end{minipage}
  \hfill
  \begin{minipage}[b]{0.49\textwidth}
    \includegraphics[width=\textwidth]{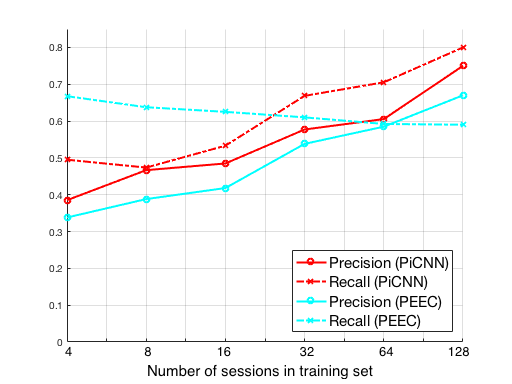}
    \caption{Effect of training set size.\label{fig:train_size}}
  \end{minipage}
\end{figure}

All four eye contact detection methods output a confidence score between 0 and 1 at each frame, with a higher score indicating increased likelihood of eye contact. We use these per-frame scores to evaluate how well the model is detecting eye contact. Since our dataset has more than 92\% negative samples, accuracy ($\frac{tp + tn}{p + n}$) is not a good measure of performance. For example, simply predicting everything as negative will give 92\% accuracy but that is not a useful detector. Instead, we report the detector's performance with respect to three metrics, namely $F_{1}$, Matthews correlation coefficient (MCC), and the Area Under Curve of Precision Recall curve (AUC-PR). 

The $F_{1}$ score is defined as
\begin{equation}
F_{1} = 2 \cdot \frac{precision \cdot recall}{precision + recall}
\end{equation}
It can be interpreted as a weighted average of the precision and recall, where 1 is the best value and 0 is the worst. In our analysis, we report our overall precision and recall at the threshold value that maximizes the $F_{1}$ score.

MCC is defined as
\begin{equation}
MCC = \frac{tp \cdot tn - fp \cdot fn}{\sqrt{(tp + fp)(tp +fn)(tn + fp)(tn + fn)}}
\end{equation}

An MCC of $1$ represents a perfect prediction, $0$ is no better than random prediction and $-1$ indicates total disagreement between prediction and ground truth. 

In the equations above, $precision = \frac{tp}{tp + fp}, recall = \frac{tp}{tp + fn}$, $tp$ is the number true positive samples, $fp$ is the number of false positive samples, $tn$ is the number true negative samples, and $fn$ is the number of false negative samples. 

Both $F_{1}$ and MCC are widely used in machine learning as a measure of the quality of binary classifiers. Because the prediction output of the models we used in our analysis is a real-numbered value instead of a hard binary output, we calculate the $F_{1}$ and MCC scores at every cutoff points and report the maximum. Additionally, we also compute the Area Under Curve of Precision Recall curve (AUC-PR). Typically, AUC is computed using Receiver Operating Characteristic (ROC) instead of Precision and Recall curve (PR curve), but AUC of ROC is not sensitive to the uneven class sizes, thus we use the PR curve. Like $F_{1}$ score, AUC-PR is maximum at 1 and minimum at 0, but it is an aggregated measure across all prediction cutoffs. 
Our final results by these criteria are summarized in Table~\ref{tab:final_scores}. Clearly, our PiCNN method outperforms all other methods when evaluated on all datasets as well as on individual groups with different conditions. 
When the results are broken down into different diagnostic groups, play protocols, gender and ethnicity, in all cases and in all metrics, our PiCNN method achieves the best performance, followed by AlexNet, PEEC and GazeLocking in this order. The standard deviation of the scores is greatest under different play protocol settings ($F_{1}$=0.0465, MCC=0.0492, AUC-PR=0.065), and second greatest under different ethnic groups ($F_{1}$=0.034, MCC=0.035, AUC-PR=0.0336). 

If smaller number of training sessions were used in a train-test split, precision tends to increase proportionally to training set size both in PEEC and PiCNN with PiCNN always performing better. However, PEEC recall is higher with smaller training set. This difference in recall is reversed after around 30 sessions and becomes more evident as more training samples are used (Fig~\ref{fig:train_size}). 
If one wants to train PiCNN model from scratch, they would need to use at least 30 sessions (100K annotated frames) to take advantage of it. If they want to include their own dataset in our framework, it is more advisable to fine-tune our pre-trained model that already has learned from more than 128 sessions.

\begin{figure}[ht]
  \centering
    \begin{subfigure}[t]{0.24\textwidth}
        \includegraphics[width=1\textwidth]{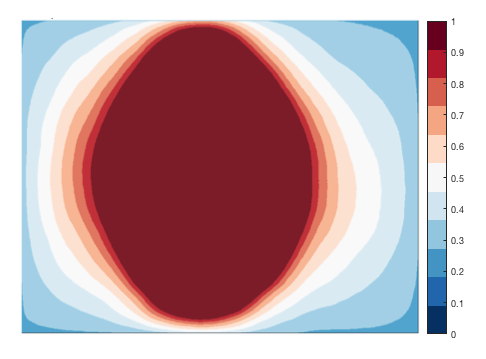}
        \caption{Face position.}
    \end{subfigure}
    \begin{subfigure}[t]{0.24\textwidth}
        \includegraphics[width=1\textwidth]{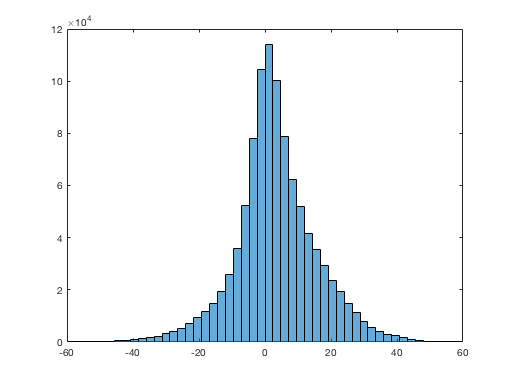}
        \caption{Yaw angle.}
    \end{subfigure}
    \begin{subfigure}[t]{0.24\textwidth}
        \includegraphics[width=1\textwidth]{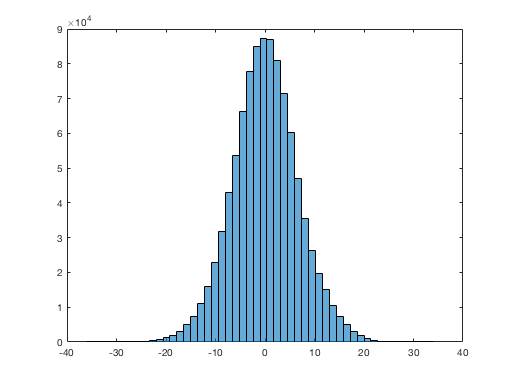}
        \caption{Pitch angle.}
    \end{subfigure}
    \begin{subfigure}[t]{0.24\textwidth}
        \includegraphics[width=1\textwidth]{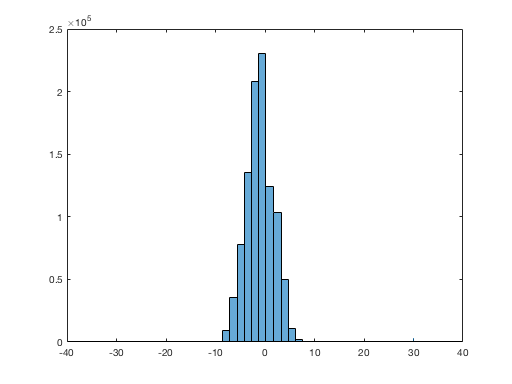}
        \caption{Roll angle.}
    \end{subfigure}    
  \caption{\textbf{Our dataset statistics.} (a) distribution of child's face location in the image shows that captured faces making eye contact often appear around image center with bigger mass vertically. (b)--(d) estimated head pose distribution in degrees.}\label{fig:dataset_dist}
\end{figure}

Fig~\ref{fig:dataset_dist} illustrates the ranges of face that our dataset covers. Although these are not a perfect representation of the data distribution because position is from the face detector's measure and pose is from the landmark detector's output, it gives an idea about the rough scope of what the current training set deals with. In case others want to apply the model to a very different setup (e.g., subject and examiner are physically configured substantially different), it might be better to do fine-tune with their own data rather than using the model as-is. Such process could make the model additionally learn unfamiliar facial features associated with eye contact, but slight variation in optical parameters (e.g., noise, exposure, motion blur, etc) would not likely need the same process because of the data augmentation step during training.

\subsubsection{Deep Visualization}
\begin{figure}[ht] 
  \centering
    \includegraphics[width=0.08\textwidth]{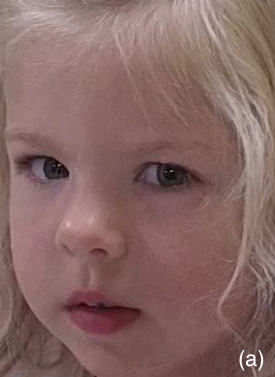}
    \includegraphics[width=0.12\textwidth]{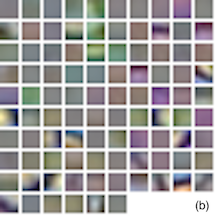}
    \includegraphics[width=0.22\textwidth]{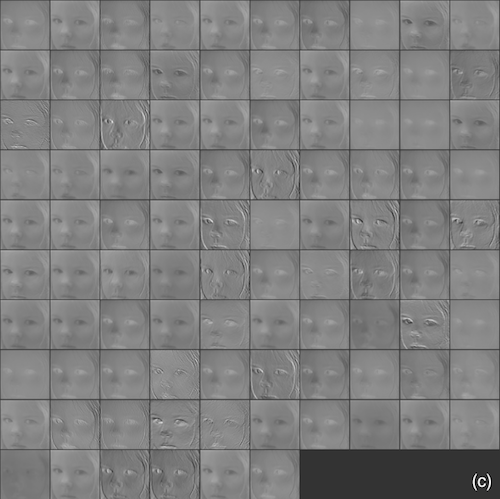}
    \includegraphics[width=0.22\textwidth]{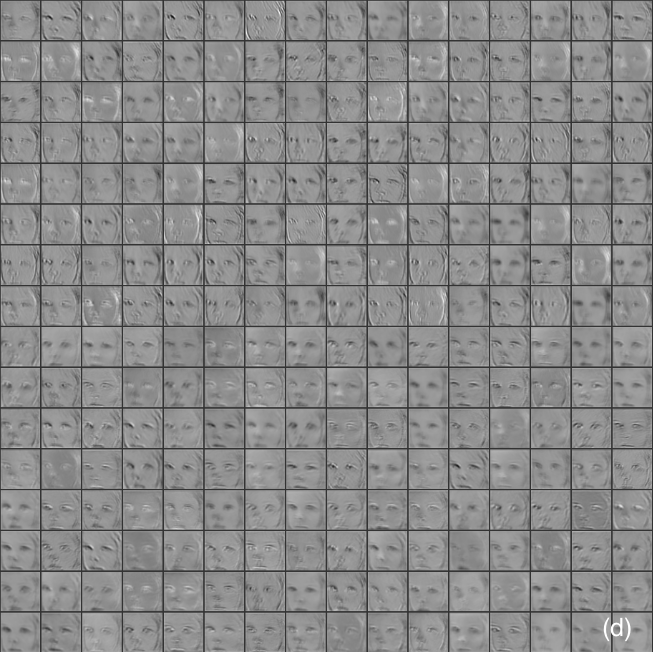}
    \includegraphics[width=0.22\textwidth]{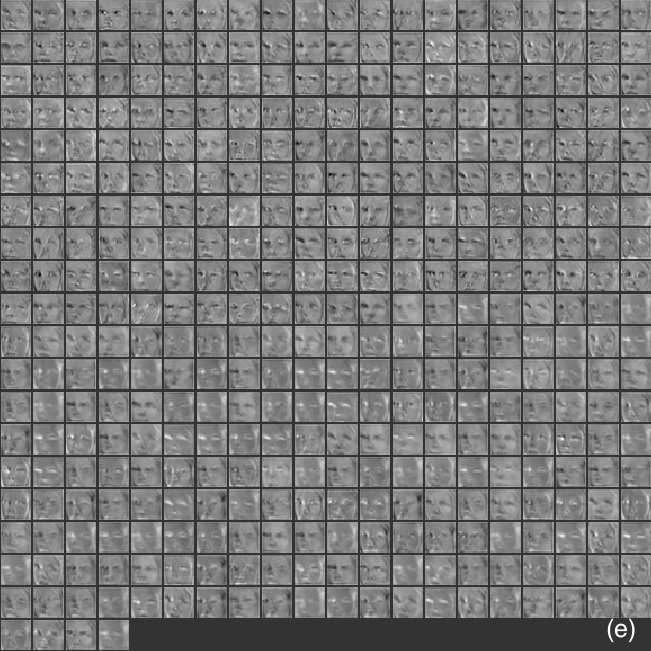}
  \caption{\textbf{Visualization of PiCNN.} (a) example input, (b) conv1 filters, (c)--(e) response map of the first 3 layers.} \label{fig:filtervis}
\end{figure}

Lastly, we visualize some of the weights of our PiCNN model to qualitatively observe what the model has learned. In Fig~\ref{fig:filtervis}, the child face image (a) is an example input, (b) shows the 96 filters learned in the first convolutional layer, and (c)--(e) are the activation maps from the first three convolutional layers when the input image is convolved with their corresponding filters. Although is not entirely clear how we can assess the usefulness of the learned weights, it is recognized that the networks have learned low-level color and round edge filters that seem useful for facial feature extraction. As shown in the activation maps, high response is produced on the regions that are valuable for eye contact detection such as the eyes and face contour.
\begin{table}[h!]
\centering
\resizebox{0.58\textwidth}{!}{%
\begin{tabular}{lllllll}
                     &                              					& $F_{1}$ & MCC & AUC-PR & Precision & Recall \\
\multirow{4}{*}{All} & PiCNN (Ours)     				&  \textbf{0.78}  &  \textbf{0.77}   &   \textbf{0.79}  &    \textbf{0.75}       &     \textbf{0.80}   \\
                     & AlexNet~\cite{krizhevsky2012imagenet}  	&  0.73  &  0.72   &   0.75  &    0.71       &     0.77   \\
                     & PEEC~\cite{ye2015detecting} 			&  0.63  &  0.62   &   0.57  &    0.67       &     0.59    \\
                     & GazeLocking~\cite{smith2013gaze} 		&  0.52  &  0.50   &   0.48  &    0.52       &     0.52    \\ \hline
\multirow{4}{*}{ASD} & PiCNN (Ours)     				&  \textbf{0.76}  &  \textbf{0.75}   &   \textbf{0.78}  &    \textbf{0.75}       &     \textbf{0.78}   \\
                     & AlexNet~\cite{krizhevsky2012imagenet}	&  0.72  &  0.71   &   0.74  &    0.69       &     0.74   \\
                     & PEEC~\cite{ye2015detecting} 			&  0.64  &  0.64   &   0.60  &    0.68       &     0.61   \\
                     & GazeLocking~\cite{smith2013gaze} 		&  0.51  &  0.49   &   0.49  &    0.50       &     0.52    \\ 
\multirow{4}{*}{TD}  & PiCNN (Ours)     				&  \textbf{0.79}  &  \textbf{0.78}   &   \textbf{0.78}  &    \textbf{0.74}       &     \textbf{0.83}   \\
                     & AlexNet~\cite{krizhevsky2012imagenet}	&  0.75  &  0.74   &   0.74  &    0.69       &     0.81   \\
                     & PEEC~\cite{ye2015detecting} 			&  0.63  &  0.62   &   0.55  &    0.65       &     0.62   \\ 
                     & GazeLocking~\cite{smith2013gaze} 		&  0.54  &  0.53   &   0.49  &    0.54       &     0.54    \\ \hline
\multirow{4}{*}{ESCS} & PiCNN (Ours)     				&  \textbf{0.75}  &  \textbf{0.75}   &   \textbf{0.76}  &    \textbf{0.69}       &     \textbf{0.82}   \\
                     & AlexNet~\cite{krizhevsky2012imagenet}	&  0.69  &  0.69   &   0.72  &    0.62       &     0.78   \\
                     & PEEC~\cite{ye2015detecting} 			&  0.57  &  0.56   &   0.47  &    0.55       &     0.60   \\
                     & GazeLocking~\cite{smith2013gaze} 		&  0.48  &  0.46   &   0.41  &    0.46       &     0.50    \\ 
\multirow{4}{*}{v-BOSCC}  & PiCNN (Ours)     			&  \textbf{0.76}  &  \textbf{0.74}   &   \textbf{0.79}  &    \textbf{0.75}       &     \textbf{0.77}   \\
                     & AlexNet~\cite{krizhevsky2012imagenet}	&  0.73  &  0.71   &   0.75  &    0.72       &     0.74   \\
                     & PEEC~\cite{ye2015detecting} 			&  0.69  &  0.67   &   0.67  &    0.72       &     0.65  \\
                     & GazeLocking~\cite{smith2013gaze} 		&  0.56  &  0.53   &   0.56  &    0.53       &     0.60    \\ 
\multirow{4}{*}{nv-BOSCC}  & PiCNN (Ours)     			&  \textbf{0.82}  &  \textbf{0.81}   &   \textbf{0.83}  &    \textbf{0.82}       &     \textbf{0.81}   \\
                     & AlexNet~\cite{krizhevsky2012imagenet}	&  0.77  &  0.76   &   0.78  &    0.76       &     0.78   \\
                     & PEEC~\cite{ye2015detecting} 			&  0.71  &  0.71   &   0.68  &    0.76       &     0.67  \\
                     & GazeLocking~\cite{smith2013gaze} 		&  0.58  &  0.56   &   0.55  &    0.57       &     0.58    \\ 
\multirow{4}{*}{R-ABC}  & PiCNN (Ours)     			&  \textbf{0.77}  &  \textbf{0.77}   &   \textbf{0.71}  &    \textbf{0.72}       &     \textbf{0.84}   \\
                     & AlexNet~\cite{krizhevsky2012imagenet} 	&  0.72  &  0.72   &   0.68  &    0.66       &     0.80   \\
                     & PEEC~\cite{ye2015detecting} 			&  0.59  &  0.59   &   0.49  &    0.66       &     0.54  \\
                     & GazeLocking~\cite{smith2013gaze} 		&  0.52  &  0.51   &   0.43  &    0.53       &     0.51    \\ 
\multirow{4}{*}{Marcus}  & PiCNN (Ours)     			&  \textbf{0.86}  &  \textbf{0.86}   &   \textbf{0.88}  &    \textbf{0.86}       &     \textbf{0.87}   \\
                     & AlexNet~\cite{krizhevsky2012imagenet}	&  0.80  &  0.79   &   0.83  &    0.77       &     0.83   \\
                     & PEEC~\cite{ye2015detecting} 			&  0.54  &  0.54   &   0.49  &    0.67       &     0.45  \\ 
                     & GazeLocking~\cite{smith2013gaze} 		&  0.46  &  0.44   &   0.41  &    0.49       &     0.44    \\ \hline
\multirow{4}{*}{male} & PiCNN (Ours)     				&  \textbf{0.77}  &  \textbf{0.75}   &   \textbf{0.77}  &    \textbf{0.74}       &     \textbf{0.79}   \\
                     & AlexNet~\cite{krizhevsky2012imagenet}	&  0.72  &  0.71   &   0.73  &    0.69       &     0.76   \\
                     & PEEC~\cite{ye2015detecting} 			&  0.64  &  0.63   &   0.57  &    0.66       &     0.61   \\
                     & GazeLocking~\cite{smith2013gaze} 		&  0.52  &  0.50   &   0.48  &    0.51       &     0.53    \\ 
\multirow{4}{*}{female}  & PiCNN (Ours)     			&  \textbf{0.78}  &  \textbf{0.78}   &   \textbf{0.80}  &    \textbf{0.76}       &     \textbf{0.81}   \\
                     & AlexNet~\cite{krizhevsky2012imagenet} 	&  0.74  &  0.74   &   0.76  &    0.70       &     0.80   \\ 
                     & PEEC~\cite{ye2015detecting} 			&  0.64  &  0.63   &   0.59  &    0.68       &     0.60   \\
                     & GazeLocking~\cite{smith2013gaze} 		&  0.53  &  0.52   &   0.49  &    0.54       &     0.52    \\ \hline
\multirow{4}{*}{Caucasian} & PiCNN (Ours)     			&  \textbf{0.76}  &  \textbf{0.75}   &   \textbf{0.76}  &    \textbf{0.73}       &     \textbf{0.79}   \\
                     & AlexNet~\cite{krizhevsky2012imagenet} 	&  0.72  &  0.71   &   0.72  &    0.68       &     0.76   \\
                     & PEEC~\cite{ye2015detecting} 			&  0.64  &  0.63   &   0.57  &    0.66       &     0.62   \\
                     & GazeLocking~\cite{smith2013gaze} 		&  0.51  &  0.50   &   0.47  &    0.51       &     0.52    \\ 
\multirow{4}{*}{mixed} & PiCNN (Ours)     				&  \textbf{0.76}  &  \textbf{0.75}   &   \textbf{0.81}  &    \textbf{0.74}       &     \textbf{0.79}   \\
                     & AlexNet~\cite{krizhevsky2012imagenet}	&  0.75  &  0.74   &   0.76  &    0.72       &     0.79   \\
                     & PEEC~\cite{ye2015detecting} 			&  0.69  &  0.67   &   0.66  &    0.70       &     0.67   \\
                     & GazeLocking~\cite{smith2013gaze} 		&  0.60  &  0.58   &   0.59  &    0.57       &     0.63    \\ 
\multirow{4}{*}{African American} & PiCNN (Ours)     		&  \textbf{0.82}  &  \textbf{0.82}   &   \textbf{0.83}  &    \textbf{0.81}       &     \textbf{0.84}   \\
                     & AlexNet~\cite{krizhevsky2012imagenet}	&  0.79  &  0.79   &   0.79  &    0.76       &     0.83   \\
                     & PEEC~\cite{ye2015detecting} 			&  0.62  &  0.61   &   0.57  &    0.68       &     0.56   \\
                     & GazeLocking~\cite{smith2013gaze} 		&  0.51  &  0.50   &   0.48  &    0.51       &     0.52    \\ 
\multirow{4}{*}{Asian} & PiCNN (Ours)     				&  \textbf{0.79}  &  \textbf{0.78}   &   \textbf{0.75}  &    \textbf{0.72}       &     \textbf{0.86}    \\
                     & AlexNet~\cite{krizhevsky2012imagenet}	&  0.67  &  0.66   &   0.63  &    0.61       &     0.74   \\
                     & PEEC~\cite{ye2015detecting} 			&  0.65  &  0.65   &   0.58  &    0.66       &     0.65   \\
                     & GazeLocking~\cite{smith2013gaze} 		&  0.49  &  0.49   &   0.41  &    0.43       &     0.57    \\ 
\multirow{4}{*}{Hispanic} & PiCNN (Ours)     			&  \textbf{0.73}  &  \textbf{0.73}   &   \textbf{0.78}  &    \textbf{0.68}       &     \textbf{0.79}   \\
                     & AlexNet~\cite{krizhevsky2012imagenet}	&  0.70  &  0.70   &   0.72  &    0.66       &     0.75   \\
                     & PEEC~\cite{ye2015detecting} 			&  0.70  &  0.69   &   0.67  &    0.69       &     0.71   \\
                     & GazeLocking~\cite{smith2013gaze} 		&  0.58  &  0.57   &   0.58  &    0.55       &     0.61    \\ 
\end{tabular}%
}
 \caption{Result scores in five metrics, grouped by different conditions. \label{tab:final_scores}}
\end{table}

\clearpage

\section{Conclusion}
In this work, we have proposed a novel approach that measures eye contact in naturalistic social interactions. We have presented a fully automated deep learning system that detects moments of eye contact from egocentric videos of adult-child interactions while also implicitly estimating the head pose. We performed a thorough and systematic evaluation of our methods on the largest ever dataset of naturalistic social interactions comprising of 22 hours of play session videos of 100 individual children. We found that our results are a significant improvement over other existing methods.

Our approach will be instrumental to understanding atypical gaze behavior in natural social settings, as well as improving the diagnosis, screening and treatment of Autism Spectrum Disorder. We also expect our work to have applications in other domains such as in the automatic analysis of turn-taking and social roles in group interactions or in the development of models of social intelligence for robots, which could allow them to interact naturally with humans based on eye contact.

\begin{acks}
%\grantsponsor{ID}{name}{url}
%\grantnum[url]{ID}{number}
The authors would like to thank Yin Li for helpful discussions, and Marcus Autism Center for sharing their dataset with the authors. 
This study was funded in part by the~\grantsponsor{11}{Simons Foundation}{} under grants~\grantnum{336363}{336363} and~\grantnum{383667}{383667}, as well as the~\grantsponsor{22}{National Science Foundation}{} under grant~\grantnum{IIS-1029679}{IIS-1029679}.

\end{acks}

%\clearpage

% Bibliography
\bibliographystyle{ACM-Reference-Format}
\bibliography{bibliography}


\begin{thebibliography}{00}

%%% the Web address.

\ifx \showCODEN    \undefined \def \showCODEN     #1{\unskip}     \fi
\ifx \showDOI      \undefined \def \showDOI       #1{{\tt DOI:}\penalty0{#1}\ }
  \fi
\ifx \showISBNx    \undefined \def \showISBNx     #1{\unskip}     \fi
\ifx \showISBNxiii \undefined \def \showISBNxiii  #1{\unskip}     \fi
\ifx \showISSN     \undefined \def \showISSN      #1{\unskip}     \fi
\ifx \showLCCN     \undefined \def \showLCCN      #1{\unskip}     \fi
\ifx \shownote     \undefined \def \shownote      #1{#1}          \fi
\ifx \showarticletitle \undefined \def \showarticletitle #1{#1}   \fi
\ifx \showURL      \undefined \def \showURL       {\relax}        \fi
% invisible to TeX
\providecommand\bibfield[2]{#2}
\providecommand\bibinfo[2]{#2}
\providecommand\natexlab[1]{#1}
\providecommand\showeprint[2][]{arXiv:#2}

\bibitem[\protect\citeauthoryear{Argyle and Dean}{Argyle and Dean}{1965}]%
        {argyle1965eye}
\bibfield{author}{\bibinfo{person}{Michael Argyle} {and} \bibinfo{person}{Janet
  Dean}.} \bibinfo{year}{1965}\natexlab{}.
\newblock \showarticletitle{Eye-contact, distance and affiliation}.
\newblock \bibinfo{journal}{{\em Sociometry\/}} (\bibinfo{year}{1965}),
  \bibinfo{pages}{289--304}.
\newblock


\bibitem[\protect\citeauthoryear{Baltru{\v{s}}aitis, Robinson, and
  Morency}{Baltru{\v{s}}aitis et~al\mbox{.}}{2014}]%
        {baltruvsaitis2014continuous}
\bibfield{author}{\bibinfo{person}{Tadas Baltru{\v{s}}aitis},
  \bibinfo{person}{Peter Robinson}, {and} \bibinfo{person}{Louis-Philippe
  Morency}.} \bibinfo{year}{2014}\natexlab{}.
\newblock \showarticletitle{Continuous conditional neural fields for structured
  regression}. In \bibinfo{booktitle}{{\em European Conference on Computer
  Vision}}. Springer, \bibinfo{pages}{593--608}.
\newblock


\bibitem[\protect\citeauthoryear{Brazelton, Tronick, Adamson, Als, and
  Wise}{Brazelton et~al\mbox{.}}{1975}]%
        {brazelton1975early}
\bibfield{author}{\bibinfo{person}{T~Berry Brazelton}, \bibinfo{person}{Edward
  Tronick}, \bibinfo{person}{Lauren Adamson}, \bibinfo{person}{Heidelise Als},
  {and} \bibinfo{person}{Susan Wise}.} \bibinfo{year}{1975}\natexlab{}.
\newblock \showarticletitle{Early mother-infant reciprocity}.
\newblock \bibinfo{journal}{{\em Parent-Infant Interaction\/}}
  \bibinfo{volume}{3} (\bibinfo{year}{1975}), \bibinfo{pages}{137}.
\newblock


\bibitem[\protect\citeauthoryear{Breiman}{Breiman}{2001}]%
        {breiman2001random}
\bibfield{author}{\bibinfo{person}{Leo Breiman}.}
  \bibinfo{year}{2001}\natexlab{}.
\newblock \showarticletitle{Random forests}.
\newblock \bibinfo{journal}{{\em Machine Learning\/}} \bibinfo{volume}{45},
  \bibinfo{number}{1} (\bibinfo{year}{2001}), \bibinfo{pages}{5--32}.
\newblock


\bibitem[\protect\citeauthoryear{Chang and Lin}{Chang and Lin}{2011}]%
        {chang2011libsvm}
\bibfield{author}{\bibinfo{person}{Chih-Chung Chang} {and}
  \bibinfo{person}{Chih-Jen Lin}.} \bibinfo{year}{2011}\natexlab{}.
\newblock \showarticletitle{LIBSVM: a library for support vector machines}.
\newblock \bibinfo{journal}{{\em ACM Transactions on Intelligent Systems and
  Technology (TIST)\/}} \bibinfo{volume}{2}, \bibinfo{number}{3}
  (\bibinfo{year}{2011}), \bibinfo{pages}{27}.
\newblock


\bibitem[\protect\citeauthoryear{Chawarska and Shic}{Chawarska and
  Shic}{2009}]%
        {chawarska2009looking}
\bibfield{author}{\bibinfo{person}{Katarzyna Chawarska} {and}
  \bibinfo{person}{Frederick Shic}.} \bibinfo{year}{2009}\natexlab{}.
\newblock \showarticletitle{Looking but not seeing: Atypical visual scanning
  and recognition of faces in 2 and 4-year-old children with autism spectrum
  disorder}.
\newblock \bibinfo{journal}{{\em Journal of Autism and Developmental
  Disorders\/}} \bibinfo{volume}{39}, \bibinfo{number}{12}
  (\bibinfo{year}{2009}), \bibinfo{pages}{1663}.
\newblock


\bibitem[\protect\citeauthoryear{Chita-Tegmark}{Chita-Tegmark}{2016}]%
        {chita2016social}
\bibfield{author}{\bibinfo{person}{Meia Chita-Tegmark}.}
  \bibinfo{year}{2016}\natexlab{}.
\newblock \showarticletitle{Social attention in ASD: a review and meta-analysis
  of eye-tracking studies}.
\newblock \bibinfo{journal}{{\em Research in Developmental Disabilities\/}}
  \bibinfo{volume}{48} (\bibinfo{year}{2016}), \bibinfo{pages}{79--93}.
\newblock


\bibitem[\protect\citeauthoryear{Daniels and Mandell}{Daniels and
  Mandell}{2014}]%
        {daniels2014explaining}
\bibfield{author}{\bibinfo{person}{Amy~M Daniels} {and}
  \bibinfo{person}{David~S Mandell}.} \bibinfo{year}{2014}\natexlab{}.
\newblock \showarticletitle{Explaining differences in age at autism spectrum
  disorder diagnosis: A critical review}.
\newblock \bibinfo{journal}{{\em Autism\/}} \bibinfo{volume}{18},
  \bibinfo{number}{5} (\bibinfo{year}{2014}), \bibinfo{pages}{583--597}.
\newblock


\bibitem[\protect\citeauthoryear{De~la Torre, Chu, Xiong, Vicente, Ding, and
  Cohn}{De~la Torre et~al\mbox{.}}{2015}]%
        {de2015intraface}
\bibfield{author}{\bibinfo{person}{Fernando De~la Torre},
  \bibinfo{person}{Wen-Sheng Chu}, \bibinfo{person}{Xuehan Xiong},
  \bibinfo{person}{Francisco Vicente}, \bibinfo{person}{Xiaoyu Ding}, {and}
  \bibinfo{person}{Jeffrey Cohn}.} \bibinfo{year}{2015}\natexlab{}.
\newblock \showarticletitle{Intraface}. In \bibinfo{booktitle}{{\em Proceedings
  of the IEEE Conference on Automatic Face and Gesture Recognition (FG)}},
  Vol.~\bibinfo{volume}{1}. IEEE, \bibinfo{pages}{1--8}.
\newblock


\bibitem[\protect\citeauthoryear{Doll{\'a}r, Welinder, and Perona}{Doll{\'a}r
  et~al\mbox{.}}{2010}]%
        {dollar2010cascaded}
\bibfield{author}{\bibinfo{person}{Piotr Doll{\'a}r}, \bibinfo{person}{Peter
  Welinder}, {and} \bibinfo{person}{Pietro Perona}.}
  \bibinfo{year}{2010}\natexlab{}.
\newblock \showarticletitle{Cascaded pose regression}. In
  \bibinfo{booktitle}{{\em Proceedings of the IEEE Conference on Computer
  Vision and Pattern Recognition (CVPR)}}. IEEE, \bibinfo{pages}{1078--1085}.
\newblock


\bibitem[\protect\citeauthoryear{Duda, Hart, and Stork}{Duda
  et~al\mbox{.}}{2012}]%
        {duda2012pattern}
\bibfield{author}{\bibinfo{person}{Richard~O Duda}, \bibinfo{person}{Peter~E
  Hart}, {and} \bibinfo{person}{David~G Stork}.}
  \bibinfo{year}{2012}\natexlab{}.
\newblock \bibinfo{booktitle}{{\em Pattern classification}}.
\newblock \bibinfo{publisher}{John Wiley \& Sons}.
\newblock


\bibitem[\protect\citeauthoryear{Edmunds, Rozga, Li, Karp, Ibanez, Rehg, and
  Stone}{Edmunds et~al\mbox{.}}{2017}]%
        {edmunds2017brief}
\bibfield{author}{\bibinfo{person}{Sarah~R Edmunds}, \bibinfo{person}{Agata
  Rozga}, \bibinfo{person}{Yin Li}, \bibinfo{person}{Elizabeth~A Karp},
  \bibinfo{person}{Lisa~V Ibanez}, \bibinfo{person}{James~M Rehg}, {and}
  \bibinfo{person}{Wendy~L Stone}.} \bibinfo{year}{2017}\natexlab{}.
\newblock \showarticletitle{Brief Report: Using a Point-of-View Camera to
  Measure Eye Gaze in Young Children with Autism Spectrum Disorder During
  Naturalistic Social Interactions: A Pilot Study}.
\newblock \bibinfo{journal}{{\em Journal of Autism and Developmental
  Disorders\/}} (\bibinfo{year}{2017}), \bibinfo{pages}{1--7}.
\newblock


\bibitem[\protect\citeauthoryear{Everingham, Van~Gool, Williams, Winn, and
  Zisserman}{Everingham et~al\mbox{.}}{2010}]%
        {everingham2010pascal}
\bibfield{author}{\bibinfo{person}{Mark Everingham}, \bibinfo{person}{Luc
  Van~Gool}, \bibinfo{person}{Christopher~KI Williams}, \bibinfo{person}{John
  Winn}, {and} \bibinfo{person}{Andrew Zisserman}.}
  \bibinfo{year}{2010}\natexlab{}.
\newblock \showarticletitle{The pascal visual object classes (voc) challenge}.
\newblock \bibinfo{journal}{{\em International Journal of Computer Vision\/}}
  \bibinfo{volume}{88}, \bibinfo{number}{2} (\bibinfo{year}{2010}),
  \bibinfo{pages}{303--338}.
\newblock


\bibitem[\protect\citeauthoryear{Felzenszwalb, Girshick, McAllester, and
  Ramanan}{Felzenszwalb et~al\mbox{.}}{2010}]%
        {felzenszwalb2010object}
\bibfield{author}{\bibinfo{person}{Pedro~F Felzenszwalb},
  \bibinfo{person}{Ross~B Girshick}, \bibinfo{person}{David McAllester}, {and}
  \bibinfo{person}{Deva Ramanan}.} \bibinfo{year}{2010}\natexlab{}.
\newblock \showarticletitle{Object detection with discriminatively trained
  part-based models}.
\newblock \bibinfo{journal}{{\em IEEE Transactions on Pattern Analysis and
  Machine Intelligence\/}} \bibinfo{volume}{32}, \bibinfo{number}{9}
  (\bibinfo{year}{2010}), \bibinfo{pages}{1627--1645}.
\newblock


\bibitem[\protect\citeauthoryear{for Disease~Control and Prevention}{for
  Disease~Control and Prevention}{2016}]%
        {cdctable}
\bibfield{author}{\bibinfo{person}{Centers for Disease~Control} {and}
  \bibinfo{person}{Prevention}.} \bibinfo{year}{2016}\natexlab{}.
\newblock \bibinfo{title}{Summary of Autism Spectrum Disorder Prevalence
  Studies}.
\newblock
  \bibinfo{howpublished}{\url{https://www.cdc.gov/ncbddd/autism/documents/ASDPrevalenceDataTable2016.pdf}}.
    (\bibinfo{year}{2016}).
\newblock
\newblock
\shownote{Accessed: 2017-05-03.}


\bibitem[\protect\citeauthoryear{Foulsham, Walker, and Kingstone}{Foulsham
  et~al\mbox{.}}{2011}]%
        {foulsham2011and}
\bibfield{author}{\bibinfo{person}{Tom Foulsham}, \bibinfo{person}{Esther
  Walker}, {and} \bibinfo{person}{Alan Kingstone}.}
  \bibinfo{year}{2011}\natexlab{}.
\newblock \showarticletitle{The where, what and when of gaze allocation in the
  lab and the natural environment}.
\newblock \bibinfo{journal}{{\em Vision Research\/}} \bibinfo{volume}{51},
  \bibinfo{number}{17} (\bibinfo{year}{2011}), \bibinfo{pages}{1920--1931}.
\newblock


\bibitem[\protect\citeauthoryear{Goodfellow, Bengio, and Courville}{Goodfellow
  et~al\mbox{.}}{2016}]%
        {Goodfellow-et-al-2016}
\bibfield{author}{\bibinfo{person}{Ian Goodfellow}, \bibinfo{person}{Yoshua
  Bengio}, {and} \bibinfo{person}{Aaron Courville}.}
  \bibinfo{year}{2016}\natexlab{}.
\newblock \bibinfo{booktitle}{{\em Deep Learning}}.
\newblock \bibinfo{publisher}{MIT Press}.
\newblock
\newblock
\shownote{\url{http://www.deeplearningbook.org}.}


\bibitem[\protect\citeauthoryear{Grzadzinski, Carr, Colombi, McGuire, Dufek,
  Pickles, and Lord}{Grzadzinski et~al\mbox{.}}{2016}]%
        {grzadzinski2016measuring}
\bibfield{author}{\bibinfo{person}{Rebecca Grzadzinski},
  \bibinfo{person}{Themba Carr}, \bibinfo{person}{Costanza Colombi},
  \bibinfo{person}{Kelly McGuire}, \bibinfo{person}{Sarah Dufek},
  \bibinfo{person}{Andrew Pickles}, {and} \bibinfo{person}{Catherine Lord}.}
  \bibinfo{year}{2016}\natexlab{}.
\newblock \showarticletitle{Measuring changes in social communication
  behaviors: preliminary development of the Brief Observation of Social
  Communication Change (BOSCC)}.
\newblock \bibinfo{journal}{{\em Journal of Autism and Developmental
  Disorders\/}} \bibinfo{volume}{46}, \bibinfo{number}{7}
  (\bibinfo{year}{2016}), \bibinfo{pages}{2464--2479}.
\newblock


\bibitem[\protect\citeauthoryear{Grzadzinski, Martinez, Gunin, Ajodan, Kim, and
  Lord}{Grzadzinski et~al\mbox{.}}{2017}]%
        {verbalboscc}
\bibfield{author}{\bibinfo{person}{R Grzadzinski}, \bibinfo{person}{K
  Martinez}, \bibinfo{person}{G Gunin}, \bibinfo{person}{E Ajodan},
  \bibinfo{person}{S Kim}, {and} \bibinfo{person}{C Lord}.}
  \bibinfo{year}{2017}\natexlab{}.
\newblock \showarticletitle{Development of the Brief Observation of Social
  Communication Change (BOSCC) for Verbally Able Children with ASD}.
\newblock \bibinfo{journal}{{\em Biennial Meeting of the Society for Research
  on Child Development (SRCD)\/}} (\bibinfo{year}{2017}).
\newblock


\bibitem[\protect\citeauthoryear{Guillon, Hadjikhani, Baduel, and
  Rog{\'e}}{Guillon et~al\mbox{.}}{2014}]%
        {guillon2014visual}
\bibfield{author}{\bibinfo{person}{Quentin Guillon}, \bibinfo{person}{Nouchine
  Hadjikhani}, \bibinfo{person}{Sophie Baduel}, {and}
  \bibinfo{person}{Bernadette Rog{\'e}}.} \bibinfo{year}{2014}\natexlab{}.
\newblock \showarticletitle{Visual social attention in autism spectrum
  disorder: Insights from eye tracking studies}.
\newblock \bibinfo{journal}{{\em Neuroscience \& Biobehavioral Reviews\/}}
  \bibinfo{volume}{42} (\bibinfo{year}{2014}), \bibinfo{pages}{279--297}.
\newblock


\bibitem[\protect\citeauthoryear{Hansen and Ji}{Hansen and Ji}{2010}]%
        {hansen2010eye}
\bibfield{author}{\bibinfo{person}{Dan~Witzner Hansen} {and}
  \bibinfo{person}{Qiang Ji}.} \bibinfo{year}{2010}\natexlab{}.
\newblock \showarticletitle{In the eye of the beholder: A survey of models for
  eyes and gaze}.
\newblock \bibinfo{journal}{{\em IEEE Transactions on Pattern Analysis and
  Machine Intelligence\/}} \bibinfo{volume}{32}, \bibinfo{number}{3}
  (\bibinfo{year}{2010}), \bibinfo{pages}{478--500}.
\newblock


\bibitem[\protect\citeauthoryear{Hosozawa, Tanaka, Shimizu, Nakano, and
  Kitazawa}{Hosozawa et~al\mbox{.}}{2012}]%
        {hosozawa2012children}
\bibfield{author}{\bibinfo{person}{Mariko Hosozawa}, \bibinfo{person}{Kyoko
  Tanaka}, \bibinfo{person}{Toshiaki Shimizu}, \bibinfo{person}{Tamami Nakano},
  {and} \bibinfo{person}{Shigeru Kitazawa}.} \bibinfo{year}{2012}\natexlab{}.
\newblock \showarticletitle{How children with specific language impairment view
  social situations: an eye tracking study}.
\newblock \bibinfo{journal}{{\em Pediatrics\/}} \bibinfo{volume}{129},
  \bibinfo{number}{6} (\bibinfo{year}{2012}), \bibinfo{pages}{e1453--e1460}.
\newblock


\bibitem[\protect\citeauthoryear{Hutman, Chela, Gillespie-Lynch, and
  Sigman}{Hutman et~al\mbox{.}}{2012}]%
        {hutman2012selective}
\bibfield{author}{\bibinfo{person}{Ted Hutman}, \bibinfo{person}{Mandeep~K
  Chela}, \bibinfo{person}{Kristen Gillespie-Lynch}, {and}
  \bibinfo{person}{Marian Sigman}.} \bibinfo{year}{2012}\natexlab{}.
\newblock \showarticletitle{Selective visual attention at twelve months: Signs
  of autism in early social interactions}.
\newblock \bibinfo{journal}{{\em Journal of Autism and Developmental
  Disorders\/}} \bibinfo{volume}{42}, \bibinfo{number}{4}
  (\bibinfo{year}{2012}), \bibinfo{pages}{487--498}.
\newblock


\bibitem[\protect\citeauthoryear{Jain and Learned-Miller}{Jain and
  Learned-Miller}{2010}]%
        {jain2010fddb}
\bibfield{author}{\bibinfo{person}{Vidit Jain} {and} \bibinfo{person}{Erik~G
  Learned-Miller}.} \bibinfo{year}{2010}\natexlab{}.
\newblock \showarticletitle{Fddb: A benchmark for face detection in
  unconstrained settings}.
\newblock \bibinfo{journal}{{\em UMass Amherst Technical Report\/}}
  (\bibinfo{year}{2010}).
\newblock


\bibitem[\protect\citeauthoryear{Jia, Shelhamer, Donahue, Karayev, Long,
  Girshick, Guadarrama, and Darrell}{Jia et~al\mbox{.}}{2014}]%
        {jia2014caffe}
\bibfield{author}{\bibinfo{person}{Yangqing Jia}, \bibinfo{person}{Evan
  Shelhamer}, \bibinfo{person}{Jeff Donahue}, \bibinfo{person}{Sergey Karayev},
  \bibinfo{person}{Jonathan Long}, \bibinfo{person}{Ross Girshick},
  \bibinfo{person}{Sergio Guadarrama}, {and} \bibinfo{person}{Trevor Darrell}.}
  \bibinfo{year}{2014}\natexlab{}.
\newblock \showarticletitle{Caffe: Convolutional architecture for fast feature
  embedding}. In \bibinfo{booktitle}{{\em Proceedings of the ACM International
  Conference on Multimedia}}. ACM, \bibinfo{pages}{675--678}.
\newblock


\bibitem[\protect\citeauthoryear{Jiang and Learned-Miller}{Jiang and
  Learned-Miller}{2016}]%
        {jiang2016face}
\bibfield{author}{\bibinfo{person}{Huaizu Jiang} {and} \bibinfo{person}{Erik
  Learned-Miller}.} \bibinfo{year}{2016}\natexlab{}.
\newblock \showarticletitle{Face detection with the faster R-CNN}.
\newblock \bibinfo{journal}{{\em arXiv preprint arXiv:1606.03473\/}}
  (\bibinfo{year}{2016}).
\newblock


\bibitem[\protect\citeauthoryear{Jones, Carr, and Klin}{Jones
  et~al\mbox{.}}{2008}]%
        {jones2008absence}
\bibfield{author}{\bibinfo{person}{Warren Jones}, \bibinfo{person}{Katelin
  Carr}, {and} \bibinfo{person}{Ami Klin}.} \bibinfo{year}{2008}\natexlab{}.
\newblock \showarticletitle{Absence of preferential looking to the eyes of
  approaching adults predicts level of social disability in 2-year-old toddlers
  with autism spectrum disorder}.
\newblock \bibinfo{journal}{{\em Archives of General Psychiatry\/}}
  \bibinfo{volume}{65}, \bibinfo{number}{8} (\bibinfo{year}{2008}),
  \bibinfo{pages}{946--954}.
\newblock


\bibitem[\protect\citeauthoryear{Kleinke}{Kleinke}{1986}]%
        {kleinke1986gaze}
\bibfield{author}{\bibinfo{person}{Chris~L Kleinke}.}
  \bibinfo{year}{1986}\natexlab{}.
\newblock \showarticletitle{Gaze and eye contact: a research review.}
\newblock \bibinfo{journal}{{\em Psychological Bulletin\/}}
  \bibinfo{volume}{100}, \bibinfo{number}{1} (\bibinfo{year}{1986}),
  \bibinfo{pages}{78}.
\newblock


\bibitem[\protect\citeauthoryear{Klin, Jones, Schultz, Volkmar, and Cohen}{Klin
  et~al\mbox{.}}{2002}]%
        {klin2002visual}
\bibfield{author}{\bibinfo{person}{Ami Klin}, \bibinfo{person}{Warren Jones},
  \bibinfo{person}{Robert Schultz}, \bibinfo{person}{Fred Volkmar}, {and}
  \bibinfo{person}{Donald Cohen}.} \bibinfo{year}{2002}\natexlab{}.
\newblock \showarticletitle{Visual fixation patterns during viewing of
  naturalistic social situations as predictors of social competence in
  individuals with autism}.
\newblock \bibinfo{journal}{{\em Archives of General Psychiatry\/}}
  \bibinfo{volume}{59}, \bibinfo{number}{9} (\bibinfo{year}{2002}),
  \bibinfo{pages}{809--816}.
\newblock


\bibitem[\protect\citeauthoryear{Krafka, Khosla, Kellnhofer, Kannan,
  Bhandarkar, Matusik, and Torralba}{Krafka et~al\mbox{.}}{2016}]%
        {krafka2016eye}
\bibfield{author}{\bibinfo{person}{Kyle Krafka}, \bibinfo{person}{Aditya
  Khosla}, \bibinfo{person}{Petr Kellnhofer}, \bibinfo{person}{Harini Kannan},
  \bibinfo{person}{Suchendra Bhandarkar}, \bibinfo{person}{Wojciech Matusik},
  {and} \bibinfo{person}{Antonio Torralba}.} \bibinfo{year}{2016}\natexlab{}.
\newblock \showarticletitle{Eye tracking for everyone}. In
  \bibinfo{booktitle}{{\em Proceedings of the IEEE Conference on Computer
  Vision and Pattern Recognition (CVPR)}}. \bibinfo{pages}{2176--2184}.
\newblock


\bibitem[\protect\citeauthoryear{Krizhevsky, Sutskever, and Hinton}{Krizhevsky
  et~al\mbox{.}}{2012}]%
        {krizhevsky2012imagenet}
\bibfield{author}{\bibinfo{person}{Alex Krizhevsky}, \bibinfo{person}{Ilya
  Sutskever}, {and} \bibinfo{person}{Geoffrey~E Hinton}.}
  \bibinfo{year}{2012}\natexlab{}.
\newblock \showarticletitle{Imagenet classification with deep convolutional
  neural networks}. In \bibinfo{booktitle}{{\em Advances in Neural Information
  Processing Systems (NIPS)}}. \bibinfo{pages}{1097--1105}.
\newblock


\bibitem[\protect\citeauthoryear{Land and Tatler}{Land and Tatler}{2009}]%
        {land2009looking}
\bibfield{author}{\bibinfo{person}{Michael Land} {and}
  \bibinfo{person}{Benjamin Tatler}.} \bibinfo{year}{2009}\natexlab{}.
\newblock \bibinfo{booktitle}{{\em Looking and acting: vision and eye movements
  in natural behaviour}}.
\newblock \bibinfo{publisher}{Oxford University Press}.
\newblock


\bibitem[\protect\citeauthoryear{Lin, Maire, Belongie, Hays, Perona, Ramanan,
  Doll{\'a}r, and Zitnick}{Lin et~al\mbox{.}}{2014}]%
        {lin2014microsoft}
\bibfield{author}{\bibinfo{person}{Tsung-Yi Lin}, \bibinfo{person}{Michael
  Maire}, \bibinfo{person}{Serge Belongie}, \bibinfo{person}{James Hays},
  \bibinfo{person}{Pietro Perona}, \bibinfo{person}{Deva Ramanan},
  \bibinfo{person}{Piotr Doll{\'a}r}, {and} \bibinfo{person}{C~Lawrence
  Zitnick}.} \bibinfo{year}{2014}\natexlab{}.
\newblock \showarticletitle{Microsoft coco: Common objects in context}. In
  \bibinfo{booktitle}{{\em European Conference on Computer Vision}}. Springer,
  \bibinfo{pages}{740--755}.
\newblock


\bibitem[\protect\citeauthoryear{Lord, DiLavore, and Gotham}{Lord
  et~al\mbox{.}}{2012}]%
        {lord2012autism}
\bibfield{author}{\bibinfo{person}{Catherine Lord}, \bibinfo{person}{Pamela~C
  DiLavore}, {and} \bibinfo{person}{Katherine Gotham}.}
  \bibinfo{year}{2012}\natexlab{}.
\newblock \bibinfo{booktitle}{{\em Autism diagnostic observation schedule}}.
\newblock \bibinfo{publisher}{Western Psychological Services Torrance, CA}.
\newblock


\bibitem[\protect\citeauthoryear{Magrelli, Jermann, Basilio, Ansermet, Hentsch,
  Nadel, and Billard}{Magrelli et~al\mbox{.}}{2013}]%
        {magrelli2013social}
\bibfield{author}{\bibinfo{person}{Silvia Magrelli}, \bibinfo{person}{Patrick
  Jermann}, \bibinfo{person}{Noris Basilio}, \bibinfo{person}{Fran{\c{c}}ois
  Ansermet}, \bibinfo{person}{Fran{\c{c}}ois Hentsch},
  \bibinfo{person}{Jaqueline Nadel}, {and} \bibinfo{person}{Aude Billard}.}
  \bibinfo{year}{2013}\natexlab{}.
\newblock \showarticletitle{Social orienting of children with autism to facial
  expressions and speech: a study with a wearable eye-tracker in naturalistic
  settings}.
\newblock \bibinfo{journal}{{\em Frontiers in Psychology\/}}
  \bibinfo{volume}{4} (\bibinfo{year}{2013}), \bibinfo{pages}{840}.
\newblock


\bibitem[\protect\citeauthoryear{Mundy and Acra}{Mundy and Acra}{2006}]%
        {mundy2006joint}
\bibfield{author}{\bibinfo{person}{P Mundy} {and}
  \bibinfo{person}{C~Fran{\c{c}}oise Acra}.} \bibinfo{year}{2006}\natexlab{}.
\newblock \showarticletitle{Joint attention, social engagement, and the
  development of social competence}.
\newblock \bibinfo{journal}{{\em The Development of Social Engagement:
  Neurobiological Perspectives\/}} (\bibinfo{year}{2006}),
  \bibinfo{pages}{81--117}.
\newblock


\bibitem[\protect\citeauthoryear{Mundy, Delgado, Block, Venezia, Hogan, and
  Seibert}{Mundy et~al\mbox{.}}{2003}]%
        {mundy2003early}
\bibfield{author}{\bibinfo{person}{Peter Mundy}, \bibinfo{person}{Christine
  Delgado}, \bibinfo{person}{Jessica Block}, \bibinfo{person}{Meg Venezia},
  \bibinfo{person}{Anne Hogan}, {and} \bibinfo{person}{Jeffrey Seibert}.}
  \bibinfo{year}{2003}\natexlab{}.
\newblock \showarticletitle{Early social communication scales (ESCS)}.
\newblock \bibinfo{journal}{{\em Coral Gables, FL: University of Miami\/}}
  (\bibinfo{year}{2003}).
\newblock


\bibitem[\protect\citeauthoryear{Noris, Nadel, Barker, Hadjikhani, and
  Billard}{Noris et~al\mbox{.}}{2012}]%
        {noris2012investigating}
\bibfield{author}{\bibinfo{person}{Basilio Noris}, \bibinfo{person}{Jacqueline
  Nadel}, \bibinfo{person}{Mandy Barker}, \bibinfo{person}{Nouchine
  Hadjikhani}, {and} \bibinfo{person}{Aude Billard}.}
  \bibinfo{year}{2012}\natexlab{}.
\newblock \showarticletitle{Investigating gaze of children with ASD in
  naturalistic settings}.
\newblock \bibinfo{journal}{{\em PloS One\/}} \bibinfo{volume}{7},
  \bibinfo{number}{9} (\bibinfo{year}{2012}), \bibinfo{pages}{e44144}.
\newblock


\bibitem[\protect\citeauthoryear{Ousley, Arriaga, Morrier, Mathys, Allen, and
  Abowd}{Ousley et~al\mbox{.}}{2013}]%
        {ousley2013beyond}
\bibfield{author}{\bibinfo{person}{Opal Ousley}, \bibinfo{person}{Rosa
  Arriaga}, \bibinfo{person}{Michael Morrier}, \bibinfo{person}{Jennifer
  Mathys}, \bibinfo{person}{Monica Allen}, {and} \bibinfo{person}{Gregory
  Abowd}.} \bibinfo{year}{2013}\natexlab{}.
\newblock \bibinfo{booktitle}{{\em Beyond parental report: findings from the
  rapid-abc, a new 4-minute interactive autism}}.
\newblock \bibinfo{type}{{T}echnical {R}eport}. \bibinfo{institution}{Technical
  report, Georgia Institute of Technology}.
\newblock


\bibitem[\protect\citeauthoryear{Parkhi, Vedaldi, and Zisserman}{Parkhi
  et~al\mbox{.}}{2015}]%
        {parkhi2015deep}
\bibfield{author}{\bibinfo{person}{Omkar~M Parkhi}, \bibinfo{person}{Andrea
  Vedaldi}, {and} \bibinfo{person}{Andrew Zisserman}.}
  \bibinfo{year}{2015}\natexlab{}.
\newblock \showarticletitle{Deep Face Recognition.}. In
  \bibinfo{booktitle}{{\em British Machine Vision Conference}},
  Vol.~\bibinfo{volume}{1}. \bibinfo{pages}{6}.
\newblock


\bibitem[\protect\citeauthoryear{Pierce, Conant, Hazin, Stoner, and
  Desmond}{Pierce et~al\mbox{.}}{2011}]%
        {pierce2011preference}
\bibfield{author}{\bibinfo{person}{Karen Pierce}, \bibinfo{person}{David
  Conant}, \bibinfo{person}{Roxana Hazin}, \bibinfo{person}{Richard Stoner},
  {and} \bibinfo{person}{Jamie Desmond}.} \bibinfo{year}{2011}\natexlab{}.
\newblock \showarticletitle{Preference for geometric patterns early in life as
  a risk factor for autism}.
\newblock \bibinfo{journal}{{\em Archives of General Psychiatry\/}}
  \bibinfo{volume}{68}, \bibinfo{number}{1} (\bibinfo{year}{2011}),
  \bibinfo{pages}{101--109}.
\newblock


\bibitem[\protect\citeauthoryear{Rehg, Abowd, Rozga, Romero, Clements,
  Sclaroff, Essa, Ousley, Li, Kim, et~al\mbox{.}}{Rehg et~al\mbox{.}}{2013}]%
        {rehg2013decoding}
\bibfield{author}{\bibinfo{person}{James Rehg}, \bibinfo{person}{Gregory
  Abowd}, \bibinfo{person}{Agata Rozga}, \bibinfo{person}{Mario Romero},
  \bibinfo{person}{Mark Clements}, \bibinfo{person}{Stan Sclaroff},
  \bibinfo{person}{Irfan Essa}, \bibinfo{person}{O Ousley},
  \bibinfo{person}{Yin Li}, \bibinfo{person}{Chanho Kim}, {and}
  \bibinfo{person}{others}.} \bibinfo{year}{2013}\natexlab{}.
\newblock \showarticletitle{Decoding children's social behavior}. In
  \bibinfo{booktitle}{{\em Proceedings of the IEEE Conference on Computer
  Vision and Pattern Recognition (CVPR)}}. \bibinfo{pages}{3414--3421}.
\newblock


\bibitem[\protect\citeauthoryear{Ren, He, Girshick, and Sun}{Ren
  et~al\mbox{.}}{2015}]%
        {ren2015faster}
\bibfield{author}{\bibinfo{person}{Shaoqing Ren}, \bibinfo{person}{Kaiming He},
  \bibinfo{person}{Ross Girshick}, {and} \bibinfo{person}{Jian Sun}.}
  \bibinfo{year}{2015}\natexlab{}.
\newblock \showarticletitle{Faster r-cnn: Towards real-time object detection
  with region proposal networks}. In \bibinfo{booktitle}{{\em Advances in
  Neural Information Processing Systems (NIPS)}}. \bibinfo{pages}{91--99}.
\newblock


\bibitem[\protect\citeauthoryear{Rozga, Hutman, Young, Rogers, Ozonoff,
  Dapretto, and Sigman}{Rozga et~al\mbox{.}}{2011}]%
        {rozga2011behavioral}
\bibfield{author}{\bibinfo{person}{Agata Rozga}, \bibinfo{person}{Ted Hutman},
  \bibinfo{person}{Gregory~S Young}, \bibinfo{person}{Sally~J Rogers},
  \bibinfo{person}{Sally Ozonoff}, \bibinfo{person}{Mirella Dapretto}, {and}
  \bibinfo{person}{Marian Sigman}.} \bibinfo{year}{2011}\natexlab{}.
\newblock \showarticletitle{Behavioral profiles of affected and unaffected
  siblings of children with autism: Contribution of measures of mother--infant
  interaction and nonverbal communication}.
\newblock \bibinfo{journal}{{\em Journal of Autism and Developmental
  Disorders\/}} \bibinfo{volume}{41}, \bibinfo{number}{3}
  (\bibinfo{year}{2011}), \bibinfo{pages}{287--301}.
\newblock


\bibitem[\protect\citeauthoryear{Rutter, Le~Couteur, and Lord}{Rutter
  et~al\mbox{.}}{2003}]%
        {rutter2003autism}
\bibfield{author}{\bibinfo{person}{Michael Rutter}, \bibinfo{person}{A
  Le~Couteur}, {and} \bibinfo{person}{C Lord}.}
  \bibinfo{year}{2003}\natexlab{}.
\newblock \showarticletitle{Autism diagnostic interview-revised}.
\newblock \bibinfo{journal}{{\em Los Angeles, CA: Western Psychological
  Services\/}}  \bibinfo{volume}{29} (\bibinfo{year}{2003}),
  \bibinfo{pages}{30}.
\newblock


\bibitem[\protect\citeauthoryear{Sasson and Elison}{Sasson and Elison}{2012}]%
        {sasson2012eye}
\bibfield{author}{\bibinfo{person}{Noah~J Sasson} {and} \bibinfo{person}{Jed~T
  Elison}.} \bibinfo{year}{2012}\natexlab{}.
\newblock \showarticletitle{Eye tracking young children with autism}.
\newblock \bibinfo{journal}{{\em Journal of Visualized Experiments\/}}
  \bibinfo{number}{61} (\bibinfo{year}{2012}), \bibinfo{pages}{e3675--e3675}.
\newblock


\bibitem[\protect\citeauthoryear{Shell, Vertegaal, Cheng, Skaburskis, Sohn,
  Stewart, Aoudeh, and Dickie}{Shell et~al\mbox{.}}{2004}]%
        {shell2004ecsglasses}
\bibfield{author}{\bibinfo{person}{Jeffrey~S Shell}, \bibinfo{person}{Roel
  Vertegaal}, \bibinfo{person}{Daniel Cheng}, \bibinfo{person}{Alexander~W
  Skaburskis}, \bibinfo{person}{Changuk Sohn}, \bibinfo{person}{A~James
  Stewart}, \bibinfo{person}{Omar Aoudeh}, {and} \bibinfo{person}{Connor
  Dickie}.} \bibinfo{year}{2004}\natexlab{}.
\newblock \showarticletitle{ECSGlasses and EyePliances: using attention to open
  sociable windows of interaction}. In \bibinfo{booktitle}{{\em Proceedings of
  the 2004 Symposium on Eye Tracking Research \& Applications (ETRA)}}. ACM,
  \bibinfo{pages}{93--100}.
\newblock


\bibitem[\protect\citeauthoryear{Sigman}{Sigman}{1998}]%
        {sigman1998emanuel}
\bibfield{author}{\bibinfo{person}{Marian Sigman}.}
  \bibinfo{year}{1998}\natexlab{}.
\newblock \showarticletitle{The Emanuel Miller Memorial Lecture 1997: Change
  and continuity in the development of children with autism}.
\newblock \bibinfo{journal}{{\em Journal of Child Psychology and Psychiatry\/}}
  \bibinfo{volume}{39}, \bibinfo{number}{6} (\bibinfo{year}{1998}),
  \bibinfo{pages}{817--827}.
\newblock


\bibitem[\protect\citeauthoryear{Sigman, Mundy, Sherman, and Ungerer}{Sigman
  et~al\mbox{.}}{1986}]%
        {sigman1986social}
\bibfield{author}{\bibinfo{person}{Marian Sigman}, \bibinfo{person}{Peter
  Mundy}, \bibinfo{person}{Tracy Sherman}, {and} \bibinfo{person}{Judy
  Ungerer}.} \bibinfo{year}{1986}\natexlab{}.
\newblock \showarticletitle{Social interactions of autistic, mentally retarded
  and normal children and their caregivers}.
\newblock \bibinfo{journal}{{\em Journal of Child Psychology and Psychiatry\/}}
  \bibinfo{volume}{27}, \bibinfo{number}{5} (\bibinfo{year}{1986}),
  \bibinfo{pages}{647--656}.
\newblock


\bibitem[\protect\citeauthoryear{Smith, Yin, Feiner, and Nayar}{Smith
  et~al\mbox{.}}{2013}]%
        {smith2013gaze}
\bibfield{author}{\bibinfo{person}{Brian~A Smith}, \bibinfo{person}{Qi Yin},
  \bibinfo{person}{Steven~K Feiner}, {and} \bibinfo{person}{Shree~K Nayar}.}
  \bibinfo{year}{2013}\natexlab{}.
\newblock \showarticletitle{Gaze locking: passive eye contact detection for
  human-object interaction}. In \bibinfo{booktitle}{{\em Proceedings of the
  26th Annual ACM Symposium on User Interface Software and Technology (UIST)}}.
  ACM, \bibinfo{pages}{271--280}.
\newblock


\bibitem[\protect\citeauthoryear{Sugano, Matsushita, and Sato}{Sugano
  et~al\mbox{.}}{2014}]%
        {sugano2014learning}
\bibfield{author}{\bibinfo{person}{Yusuke Sugano}, \bibinfo{person}{Yasuyuki
  Matsushita}, {and} \bibinfo{person}{Yoichi Sato}.}
  \bibinfo{year}{2014}\natexlab{}.
\newblock \showarticletitle{Learning-by-synthesis for appearance-based 3d gaze
  estimation}. In \bibinfo{booktitle}{{\em Proceedings of the IEEE Conference
  on Computer Vision and Pattern Recognition (CVPR)}}.
  \bibinfo{pages}{1821--1828}.
\newblock


\bibitem[\protect\citeauthoryear{Turk and Pentland}{Turk and Pentland}{1991}]%
        {turk1991face}
\bibfield{author}{\bibinfo{person}{Matthew~A Turk} {and}
  \bibinfo{person}{Alex~P Pentland}.} \bibinfo{year}{1991}\natexlab{}.
\newblock \showarticletitle{Face recognition using eigenfaces}. In
  \bibinfo{booktitle}{{\em Proceedings of the IEEE Conference on Computer
  Vision and Pattern Recognition (CVPR)}}. IEEE, \bibinfo{pages}{586--591}.
\newblock


\bibitem[\protect\citeauthoryear{Viola and Jones}{Viola and Jones}{2004}]%
        {viola2004robust}
\bibfield{author}{\bibinfo{person}{Paul Viola} {and} \bibinfo{person}{Michael~J
  Jones}.} \bibinfo{year}{2004}\natexlab{}.
\newblock \showarticletitle{Robust real-time face detection}.
\newblock \bibinfo{journal}{{\em International Journal of Computer Vision\/}}
  \bibinfo{volume}{57}, \bibinfo{number}{2} (\bibinfo{year}{2004}),
  \bibinfo{pages}{137--154}.
\newblock


\bibitem[\protect\citeauthoryear{vision}{vision}{2017}]%
        {omron}
\bibfield{author}{\bibinfo{person}{OMRON~OKAO vision}.}
  \bibinfo{year}{2017}\natexlab{}.
\newblock
  \bibinfo{howpublished}{\url{https://www.omron.com/ecb/products/mobile/okao01.html}}.
    (\bibinfo{year}{2017}).
\newblock
\newblock
\shownote{Accessed: 2017-05-03.}


\bibitem[\protect\citeauthoryear{Yang, Luo, Loy, and Tang}{Yang
  et~al\mbox{.}}{2015}]%
        {yang2015facial}
\bibfield{author}{\bibinfo{person}{Shuo Yang}, \bibinfo{person}{Ping Luo},
  \bibinfo{person}{Chen-Change Loy}, {and} \bibinfo{person}{Xiaoou Tang}.}
  \bibinfo{year}{2015}\natexlab{}.
\newblock \showarticletitle{From facial parts responses to face detection: A
  deep learning approach}. In \bibinfo{booktitle}{{\em Proceedings of the IEEE
  International Conference on Computer Vision (ICCV)}}.
  \bibinfo{pages}{3676--3684}.
\newblock


\bibitem[\protect\citeauthoryear{Yang, Luo, Loy, and Tang}{Yang
  et~al\mbox{.}}{2016}]%
        {yang2016wider}
\bibfield{author}{\bibinfo{person}{Shuo Yang}, \bibinfo{person}{Ping Luo},
  \bibinfo{person}{Chen~Change Loy}, {and} \bibinfo{person}{Xiaoou Tang}.}
  \bibinfo{year}{2016}\natexlab{}.
\newblock \showarticletitle{WIDER FACE: A Face Detection Benchmark}. In
  \bibinfo{booktitle}{{\em IEEE Conference on Computer Vision and Pattern
  Recognition (CVPR)}}.
\newblock


\bibitem[\protect\citeauthoryear{Ye, Li, Fathi, Han, Rozga, Abowd, and Rehg}{Ye
  et~al\mbox{.}}{2012}]%
        {ye2012detecting}
\bibfield{author}{\bibinfo{person}{Zhefan Ye}, \bibinfo{person}{Yin Li},
  \bibinfo{person}{Alireza Fathi}, \bibinfo{person}{Yi Han},
  \bibinfo{person}{Agata Rozga}, \bibinfo{person}{Gregory~D Abowd}, {and}
  \bibinfo{person}{James~M Rehg}.} \bibinfo{year}{2012}\natexlab{}.
\newblock \showarticletitle{Detecting eye contact using wearable eye-tracking
  glasses}. In \bibinfo{booktitle}{{\em Proceedings of the ACM Conference on
  Ubiquitous Computing}}. ACM, \bibinfo{pages}{699--704}.
\newblock


\bibitem[\protect\citeauthoryear{Ye, Li, Liu, Bridges, Rozga, and Rehg}{Ye
  et~al\mbox{.}}{2015}]%
        {ye2015detecting}
\bibfield{author}{\bibinfo{person}{Zhefan Ye}, \bibinfo{person}{Yin Li},
  \bibinfo{person}{Yun Liu}, \bibinfo{person}{Chanel Bridges},
  \bibinfo{person}{Agata Rozga}, {and} \bibinfo{person}{James~M Rehg}.}
  \bibinfo{year}{2015}\natexlab{}.
\newblock \showarticletitle{Detecting bids for eye contact using a wearable
  camera}. In \bibinfo{booktitle}{{\em Proceedings of the IEEE Conference on
  Automatic Face and Gesture Recognition (FG)}}, Vol.~\bibinfo{volume}{1}.
  IEEE, \bibinfo{pages}{1--8}.
\newblock


\bibitem[\protect\citeauthoryear{Zhang, Zhang, Li, and Qiao}{Zhang
  et~al\mbox{.}}{2016b}]%
        {zhang2016joint}
\bibfield{author}{\bibinfo{person}{Kaipeng Zhang}, \bibinfo{person}{Zhanpeng
  Zhang}, \bibinfo{person}{Zhifeng Li}, {and} \bibinfo{person}{Yu Qiao}.}
  \bibinfo{year}{2016}\natexlab{b}.
\newblock \showarticletitle{Joint Face Detection and Alignment Using Multitask
  Cascaded Convolutional Networks}.
\newblock \bibinfo{journal}{{\em IEEE Signal Processing Letters\/}}
  \bibinfo{volume}{23}, \bibinfo{number}{10} (\bibinfo{year}{2016}),
  \bibinfo{pages}{1499--1503}.
\newblock


\bibitem[\protect\citeauthoryear{Zhang, Sugano, Fritz, and Bulling}{Zhang
  et~al\mbox{.}}{2015}]%
        {zhang2015appearance}
\bibfield{author}{\bibinfo{person}{Xucong Zhang}, \bibinfo{person}{Yusuke
  Sugano}, \bibinfo{person}{Mario Fritz}, {and} \bibinfo{person}{Andreas
  Bulling}.} \bibinfo{year}{2015}\natexlab{}.
\newblock \showarticletitle{Appearance-based gaze estimation in the wild}. In
  \bibinfo{booktitle}{{\em Proceedings of the IEEE Conference on Computer
  Vision and Pattern Recognition (CVPR)}}. \bibinfo{pages}{4511--4520}.
\newblock


\bibitem[\protect\citeauthoryear{Zhang, Sugano, Fritz, and Bulling}{Zhang
  et~al\mbox{.}}{2016a}]%
        {bulling2017arxiv}
\bibfield{author}{\bibinfo{person}{Xucong Zhang}, \bibinfo{person}{Yusuke
  Sugano}, \bibinfo{person}{Mario Fritz}, {and} \bibinfo{person}{Andreas
  Bulling}.} \bibinfo{year}{2016}\natexlab{a}.
\newblock \showarticletitle{It's Written All Over Your Face: Full-Face
  Appearance-Based Gaze Estimation}.
\newblock \bibinfo{journal}{{\em arXiv preprint arXiv:1611.08860\/}}
  (\bibinfo{year}{2016}).
\newblock


\bibitem[\protect\citeauthoryear{Zwaigenbaum, Bryson, and Garon}{Zwaigenbaum
  et~al\mbox{.}}{2013}]%
        {zwaigenbaum2013early}
\bibfield{author}{\bibinfo{person}{Lonnie Zwaigenbaum}, \bibinfo{person}{Susan
  Bryson}, {and} \bibinfo{person}{Nancy Garon}.}
  \bibinfo{year}{2013}\natexlab{}.
\newblock \showarticletitle{Early identification of autism spectrum disorders}.
\newblock \bibinfo{journal}{{\em Behavioral Brain Research\/}}
  \bibinfo{volume}{251} (\bibinfo{year}{2013}), \bibinfo{pages}{133--146}.
\newblock


\end{thebibliography}

\end{document}